
\input harvmac
\input amssym



\newfam\frakfam
\font\teneufm=eufm10
\font\seveneufm=eufm7
\font\fiveeufm=eufm5
\textfont\frakfam=\teneufm
\scriptfont\frakfam=\seveneufm
\scriptscriptfont\frakfam=\fiveeufm


\def\bb{
\font\tenmsb=msbm10
\font\sevenmsb=msbm7
\font\fivemsb=msbm5
\textfont1=\tenmsb
\scriptfont1=\sevenmsb
\scriptscriptfont1=\fivemsb
}



\newfam\dsromfam
\font\tendsrom=dsrom10
\textfont\dsromfam=\tendsrom
\def\ds{\fam\dsromfam \tendsrom}


\newfam\mbffam
\font\tenmbf=cmmib10
\font\sevenmbf=cmmib7
\font\fivembf=cmmib5
\textfont\mbffam=\tenmbf
\scriptfont\mbffam=\sevenmbf
\scriptscriptfont\mbffam=\fivembf
\def\mbf{\fam\mbffam \tenmbf}


\newfam\mbfcalfam
\font\tenmbfcal=cmbsy10
\font\sevenmbfcal=cmbsy7
\font\fivembfcal=cmbsy5
\textfont\mbfcalfam=\tenmbfcal
\scriptfont\mbfcalfam=\sevenmbfcal
\scriptscriptfont\mbfcalfam=\fivembfcal


\newfam\mscrfam
\font\tenmscr=rsfs10
\font\sevenmscr=rsfs7
\font\fivemscr=rsfs5
\textfont\mscrfam=\tenmscr
\scriptfont\mscrfam=\sevenmscr
\scriptscriptfont\mscrfam=\fivemscr




\def\tilde{\widetilde}
\def\t{\tilde}
\def\hat{\widehat}

\def\bar{\overline}
\def\b{\bar}
\def\bsq#1{{{\b{#1}}^{\lower 2.5pt\hbox{$\scriptstyle 2$}}}}
\def\bexp#1#2{{{\b{#1}}^{\lower 2.5pt\hbox{$\scriptstyle #2$}}}}
\def\dotexp#1#2{{{#1}^{\lower 2.5pt\hbox{$\scriptstyle #2$}}}}


\def\rt2{\sqrt{2}}
\def\half {{1 \over 2}}
\def\Re{\mathop{\rm Re}}
\def\Im{\mathop{\rm Im}}
\def\d{\partial}

\def\grad{\nabla}

\def\det{\mathop{\rm det}}

\def\vector#1{{\mbf \vec{#1}}}


\font\tenbifull=cmmib10
\font\tenbimed=cmmib7
\font\tenbismall=cmmib5
\textfont9=\tenbifull \scriptfont9=\tenbimed
\scriptscriptfont9=\tenbismall

\mathchardef\bbGamma="7000
\mathchardef\bbDelta="7001
\mathchardef\bbPhi="7002
\mathchardef\bbAlpha="7003
\mathchardef\bbXi="7004
\mathchardef\bbPi="7005
\mathchardef\bbSigma="7006
\mathchardef\bbUpsilon="7007
\mathchardef\bbTheta="7008
\mathchardef\bbPsi="7009
\mathchardef\bbOmega="700A
\mathchardef\bbalpha="710B
\mathchardef\bbbeta="710C
\mathchardef\bbgamma="710D
\mathchardef\bbdelta="710E
\mathchardef\bbepsilon="710F
\mathchardef\bbzeta="7110
\mathchardef\bbeta="7111
\mathchardef\bbtheta="7112
\mathchardef\bbiota="7113
\mathchardef\bbkappa="7114
\mathchardef\bblambda="7115
\mathchardef\bbmu="7116
\mathchardef\bbnu="7117
\mathchardef\bbxi="7118
\mathchardef\bbpi="7119
\mathchardef\bbrho="711A
\mathchardef\bbsigma="711B
\mathchardef\bbtau="711C
\mathchardef\bbupsilon="711D
\mathchardef\bbphi="711E
\mathchardef\bbchi="711F
\mathchardef\bbpsi="7120
\mathchardef\bbomega="7121
\mathchardef\bbvarepsilon="7122
\mathchardef\bbvartheta="7123
\mathchardef\bbvarpi="7124
\mathchardef\bbvarrho="7125
\mathchardef\bbvarsigma="7126
\mathchardef\bbvarphi="7127


\def\alphadot{{\dot\alpha}}




\def\CK{{\cal K}}
\def\CL{{\cal L}}
\def\CM{{\cal M}}
\def\CN{{\cal N}}
\def\CO{{\cal O}}

\def\CR{{\cal R}}


\def\1{{\ds 1}}
\def\R{\hbox{$\bb R$}}
\def\C{\hbox{$\bb C$}}


\def\ep{\varepsilon}

\noblackbox




\lref\DumitrescuIU{
  T.~T.~Dumitrescu, N.~Seiberg,
  ``Supercurrents and Brane Currents in Diverse Dimensions,''
[arXiv:1106.0031 [hep-th]].
}

\lref\FestucciaWS{
  G.~Festuccia, N.~Seiberg,
  ``Rigid Supersymmetric Theories in Curved Superspace,''
JHEPA,1106,114.\ 2011 {\bf 1106}, 114 (2011).
[arXiv:1105.0689 [hep-th]].
}

\lref\StelleYE{
  K.~S.~Stelle, P.~C.~West,
  ``Minimal Auxiliary Fields for Supergravity,''
Phys.\ Lett.\  {\bf B74}, 330 (1978).
}

\lref\FerraraEM{
  S.~Ferrara, P.~van Nieuwenhuizen,
  ``The Auxiliary Fields of Supergravity,''
Phys.\ Lett.\  {\bf B74}, 333 (1978).
}

\lref\SohniusTP{
  M.~F.~Sohnius, P.~C.~West,
  ``An Alternative Minimal Off-Shell Version of N=1 Supergravity,''
Phys.\ Lett.\  {\bf B105}, 353 (1981).
}

\lref\SohniusFW{
  M.~Sohnius, P.~C.~West,
  ``The Tensor Calculus And Matter Coupling Of The Alternative Minimal Auxiliary Field Formulation Of N=1 Supergravity,''
Nucl.\ Phys.\  {\bf B198}, 493 (1982).
}

\lref\FerraraPZ{
  S.~Ferrara, B.~Zumino,
  ``Transformation Properties of the Supercurrent,''
Nucl.\ Phys.\  {\bf B87}, 207 (1975).
}

\lref\GatesNR{
  S.~J.~Gates, M.~T.~Grisaru, M.~Rocek, W.~Siegel,
  ``Superspace Or One Thousand and One Lessons in Supersymmetry,''
Front.\ Phys.\  {\bf 58}, 1-548 (1983).
[hep-th/0108200].
}

\lref\KomargodskiRB{
  Z.~Komargodski, N.~Seiberg,
  ``Comments on Supercurrent Multiplets, Supersymmetric Field Theories and Supergravity,''
JHEP {\bf 1007}, 017 (2010).
[arXiv:1002.2228 [hep-th]].
}

\lref\Kosmann{
  Y.~Kosmann,
  ``D\'eriv\'ees de Lie des spineurs,''
Ann.\ di Matematica Pura e Appl.\  {\bf 91}, 317395 (1972).
}

\lref\Boyertwo{
C.~P.~Boyer,
  ``A note on hyper-Hermitian four-manifolds"
Proc.\ Amer.\ Math.\ Soc.\ {\bf 102}, 157-164 (1988).
}

\lref\WittenZE{
  E.~Witten,
  ``Topological Quantum Field Theory,''
Commun.\ Math.\ Phys.\  {\bf 117}, 353 (1988).
}

\lref\SamtlebenGY{
  H.~Samtleben and D.~Tsimpis,
  ``Rigid supersymmetric theories in 4d Riemannian space,''
[arXiv:1203.3420 [hep-th]].
}

\lref\JiaHW{
  B.~Jia and E.~Sharpe,
  ``Rigidly Supersymmetric Gauge Theories on Curved Superspace,''
[arXiv:1109.5421 [hep-th]].
}

\lref\RomelsbergerEG{
  C.~Romelsberger,
  ``Counting chiral primaries in N = 1, d=4 superconformal field theories,''
Nucl.\ Phys.\ B {\bf 747}, 329 (2006).
[hep-th/0510060].
}

\lref\PestunRZ{
  V.~Pestun,
  ``Localization of gauge theory on a four-sphere and supersymmetric Wilson loops,''
Commun.\ Math.\ Phys.\  {\bf 313}, 71 (2012).
[arXiv:0712.2824 [hep-th]].
}

\lref\KapustinKZ{
  A.~Kapustin, B.~Willett and I.~Yaakov,
  ``Exact Results for Wilson Loops in Superconformal Chern-Simons Theories with Matter,''
JHEP {\bf 1003}, 089 (2010).
[arXiv:0909.4559 [hep-th]].
}

\lref\JafferisUN{
  D.~L.~Jafferis,
  ``The Exact Superconformal R-Symmetry Extremizes Z,''
JHEP {\bf 1205}, 159 (2012).
[arXiv:1012.3210 [hep-th]].
}

\lref\HamaAV{
  N.~Hama, K.~Hosomichi and S.~Lee,
  ``Notes on SUSY Gauge Theories on Three-Sphere,''
JHEP {\bf 1103}, 127 (2011).
[arXiv:1012.3512 [hep-th]].
}

\lref\LawsonYR{
  H.~B.~Lawson and M.~L.~Michelsohn,
  ``Spin geometry,''
(Princeton mathematical series. 38).
}

\lref\WessCP{
  J.~Wess, J.~Bagger,
  ``Supersymmetry and Supergravity,''
Princeton, Univ. Pr. (1992).
}

\lref\DumitrescuHE{
  T.~T.~Dumitrescu, G.~Festuccia and N.~Seiberg,
  ``Exploring Curved Superspace,''
JHEP {\bf 1208}, 141 (2012).
[arXiv:1205.1115 [hep-th]].
}

\lref\KlareGN{
  C.~Klare, A.~Tomasiello and A.~Zaffaroni,
  ``Supersymmetry on Curved Spaces and Holography,''
JHEP {\bf 1208}, 061 (2012).
[arXiv:1205.1062 [hep-th]].
}

\lref\CassaniRI{
  D.~Cassani, C.~Klare, D.~Martelli, A.~Tomasiello and A.~Zaffaroni,
  ``Supersymmetry in Lorentzian Curved Spaces and Holography,''
[arXiv:1207.2181 [hep-th]].
}

\lref\LiuBI{
  J.~T.~Liu, L.~A.~P.~Zayas and D.~Reichmann,
  ``Rigid Supersymmetric Backgrounds of Minimal Off-Shell Supergravity,''
[arXiv:1207.2785 [hep-th]].
}

\lref\ClossetVP{
  C.~Closset, T.~T.~Dumitrescu, G.~Festuccia, Z.~Komargodski and N.~Seiberg,
  ``Comments on Chern-Simons Contact Terms in Three Dimensions,''
[arXiv:1206.5218 [hep-th]].
}

\lref\ClossetVG{
  C.~Closset, T.~T.~Dumitrescu, G.~Festuccia, Z.~Komargodski and N.~Seiberg,
  ``Contact Terms, Unitarity, and F-Maximization in Three-Dimensional Superconformal Theories,''
[arXiv:1205.4142 [hep-th]].
}

\lref\Barth{
  W.~Barth, C.~Peters, A.~Van de Ven,
  ``Compact Complex Surfaces,''
  Springer (1984).
}

\lref\BuchbinderUQ{
  I.~L.~Buchbinder and S.~M.~Kuzenko,
  ``Ideas and methods of supersymmetry and supergravity: A Walk through superspace,''
Bristol, UK: IOP (1995) 640 p.
}


\rightline{PUPT-2425}
\Title{
} {\vbox{\centerline{Exploring Curved Superspace (II)}}}

\centerline{ Thomas T. Dumitrescu\hskip0.5pt$^{1}$ and Guido Festuccia\hskip0.5pt$^{2}$}
\bigskip
 \centerline{$^{1}$ {\it Department of Physics, Princeton University, Princeton, NJ 08544, USA}}
  \centerline{$^{2}${\it
Institute for Advanced Study, Princeton, NJ 08540, USA}}

\vskip50pt

\noindent We extend our previous analysis of Riemannian four-manifolds~$\CM$ admitting rigid supersymmetry to~$\CN=1$ theories that do not possess a~$U(1)_R$ symmetry. With one exception, we find that~$\CM$ must be a Hermitian manifold. However, the presence of supersymmetry imposes additional restrictions. For instance, a supercharge that squares to zero exists, if the canonical bundle of the Hermitian manifold~$\CM$ admits a nowhere vanishing, holomorphic section. This requirement can be slightly relaxed if~$\CM$ is a torus bundle over a Riemann surface, in which case we obtain a supercharge that squares to a complex Killing vector. We also analyze the conditions for the presence of more than one supercharge. The exceptional case occurs when~$\CM$ is a warped product~$S^3 \times \R$, where the radius of the round~$S^3$ is allowed to vary along~$\R$. Such manifolds admit two supercharges that generate the superalgebra~$OSp(1|2)$. If the~$S^3$ smoothly shrinks to zero at two points, we obtain a squashed four-sphere, which is not a Hermitian manifold.

\vskip35pt

\Date{September 2012}


\newsec{Introduction}

The study of supersymmetric field theories on curved manifolds leads to new observables, which can often be calculated exactly using localization techniques~\refs{\WittenZE\RomelsbergerEG\PestunRZ\KapustinKZ\JafferisUN\HamaAV\ClossetVG-\ClossetVP}. Most authors have focused on supersymmetric theories with an~$R$-symmetry. Riemannian four-manifolds that admit rigid supersymmetry for~$\CN=1$ theories with a~$U(1)_R$ symmetry were classified in~\refs{\DumitrescuHE,\KlareGN}. In this paper, we will complete the classification by analyzing theories that do not possess a~$U(1)_R$ symmetry. For recent progress in this direction, see also~\refs{\FestucciaWS\JiaHW\SamtlebenGY-\LiuBI}.

As in~\DumitrescuHE, we follow the approach developed in~\FestucciaWS, which is based on the rigid limit of off-shell supergravity. In this limit, the fields in the supergravity multiplet are taken to be arbitrary classical backgrounds. A given background configuration possesses rigid supersymmetry whenever it is invariant under some subalgebra of the local supergravity transformations. (See also~\BuchbinderUQ\ for an introduction to rigid superspace geometry.) Four-dimensional~$\CN=1$ theories can be coupled to different formulations of off-shell supergravity, depending on the structure of their supercurrent multiplet. (See~\refs{\KomargodskiRB,\DumitrescuIU} for a recent discussion of supercurrent multiplets.) Theories with a~$U(1)_R$ symmetry admit an~$\CR$-multiplet (see for instance~\GatesNR) and can be coupled to new minimal supergravity~\refs{\SohniusTP,\SohniusFW}. Such theories were analyzed in~\refs{\DumitrescuHE,\KlareGN}.

In this paper, we will consider four-dimensional~$\CN=1$ theories that admit a Ferrara-Zumino supercurrent multiplet~\FerraraPZ. These theories can be coupled to old minimal supergravity~\refs{\StelleYE,\FerraraEM}. In this formulation, the supergravity multiplet consists of the metric~$g_{\mu\nu}$ and the gravitino~$\psi_{\mu\alpha}$, as well as a vector field~$b^\mu$ and two scalar fields~$M$ and~$\t M$. It is important to emphasize that~$b^\mu$ is a globally well-defined vector field, which is not subject to any gauge redundancy. The scalars~$M$ and~$\t M$ should be thought of as the dual four-form field strengths for certain three-form gauge fields. In ordinary supergravity, $b^\mu, M$, and~$\t M$ are auxiliary fields that can be eliminated by solving their equations of motion. Here we regard them as background fields that can take arbitrary values.

In old minimal supergravity, the local supersymmetry variation of the gravitino takes the form
\eqn\gravvar{\eqalign{& \delta \psi_\mu = - 2 \grad_\mu \zeta + {i \over 3} M \sigma_\mu \t \zeta + {2 i \over 3} b_\mu \zeta + {2 i \over 3} b^\nu \sigma_{\mu\nu} \zeta~,\cr
& \delta \t \psi_\mu = -2 \grad_\mu \t \zeta + {i \over 3} \t M \t \sigma_\mu \zeta - {2 i \over 3} b_\mu \t \zeta - {2 i \over 3} b^\nu \t \sigma_{\mu\nu} \t \zeta~.}}
Here the spinor parameter~$\zeta$ is left-handed and carries un-dotted indices,~$\zeta_\alpha$, while~$\t \zeta$ is right-handed and carries dotted indices, $\t \zeta^\alphadot$, and similarly for~$\psi_\mu$ and~$\t \psi_\mu$. (Our conventions are summarized in appendix~A.) In Euclidean signature, the spinors~$\zeta$ and~$\t \zeta$ are independent and the background fields~$b^\mu, M$, and~$\t M$ are generally complex.\foot{In Lorentzian signature, $\zeta$ and~$\t \zeta$ are exchanged by complex conjugation, $b^\mu$ is real, and~$\t M$ is the complex conjugate of~$M$. Rigid supersymmetry on Lorentzian manifolds was recently discussed in~\refs{\CassaniRI,\LiuBI}.} However, we will always take the metric~$g_{\mu\nu}$ to be real.

The background fields~$g_{\mu\nu}, b^\mu, M$, and~$\t M$ preserve rigid supersymmetry, if and only if there is a supercharge~$Q = (\zeta, \t \zeta)$ for which the variations~\gravvar\ vanish,
\eqn\speq{\eqalign{& \grad_\mu \zeta = {i \over 6} M \sigma_\mu \t \zeta + {i \over 3} b_\mu \zeta + {i \over 3} b^\nu \sigma_{\mu\nu} \zeta~,\cr
& \grad_\mu \t \zeta = {i \over 6} \t M \t \sigma_\mu \zeta - {i \over 3} b_\mu \t \zeta - {i \over 3} b^\nu \t \sigma_{\mu\nu} \t \zeta~.}}
The algebra generated by~$Q$ follows from the algebra of supergravity transformations~\refs{\StelleYE,\FerraraEM},
\eqn\salg{\eqalign{&\delta_Q^2 = 2 i \delta_K~, \qquad K^\mu = \zeta \sigma^\mu \t \zeta~,\cr
& [\delta_K, \delta_Q] = 0~.}}
Here~$\delta_Q$ is the supersymmetry variation corresponding to~$Q$ and~$\delta_K = \CL_K$ is the usual Lie derivative. (The action of the Lie derivative on spinors is reviewed in appendix~A.) As we will see below, $K$ is a Killing vector. Given another supercharge~$Q' = (\eta, \t \eta)$, its (anti-)~commutation relations with~$Q$ and~$K$ are given by
\eqn\twosc{\eqalign{& \{ \delta_Q, \delta_{Q'}\} = 2 i \delta_Y~, \qquad Y^\mu = \zeta \sigma^\mu \t \eta + \eta \sigma^\mu \t \zeta~,\cr
& [\delta_K, \delta_{Q'}] = - \delta_{\CL_K Q'}~, \qquad \CL_K Q' = (\CL_K \eta, \CL_K \t \eta)~.}}
Here~$Y$ is also a Killing vector.

In this paper, we will analyze Riemannian four-manifolds~$\CM$ that admit one or several solutions of~\speq. Such solutions do not exist for arbitrary configurations of the background fields~$g_{\mu\nu}, b^\mu, M$, and~$\t M$. Locally, the equations are only consistent if the background fields satisfy certain integrability conditions. Additionally, there may be global obstructions. We would like to understand the restrictions on the background fields due to one or several solutions of~\speq. Conversely, we would like to formulate sufficient conditions for the existence of such solutions.

In section~2, we will discuss general properties of spinor pairs~$(\zeta, \t \zeta)$ that satisfy the equations~\speq, as well as various bilinears that can be constructed using~$\zeta$ and~$\t \zeta$. In particular, we investigate the conditions under which the spinors are nowhere vanishing on~$\CM$. Whenever this is the case, it follows that~$\CM$ admits an integrable complex structure that is compatible with the metric, and hence it is a Hermitian manifold.

In section~3 we consider Hermitian manifolds~$\CM$ that admit a single supercharge. Here we distinguish two cases:
\medskip
\item{1.)} If~$\zeta$ is nowhere vanishing and~$\t \zeta$ vanishes identically, then we obtain a supercharge $Q = (\zeta, 0)$ that squares to zero, $\delta_Q^2 = 0$. Such a solution exists whenever~$\CM$ admits a nowhere vanishing, anti-holomorphic section~$p$ of its anti-canonical bundle~$\b \CK = \Lambda^{0,2}$ of complex~$(0,2)$ forms. (Equivalently, a nowhere vanishing, holomorphic section of its canonical bundle~$\CK = \Lambda^{2,0}$.)

\item{2.)} If both~$\zeta$ and~$\t \zeta$ are nowhere vanishing, we obtain a supercharge~$Q = (\zeta, \t \zeta)$ that squares to a complex Killing vector~$K^\mu = \zeta \sigma^\mu \t \zeta$ as in~\salg. If~$K$ commutes with its complex conjugate~$\b K$ so that~$[K, \b K] = 0$, the Hermitian metric on~$\CM$ must take the form
\eqn\torusfib{ds^2 = \Omega(z, \b z)^2 \big( (dw  + h(z, \b z) dz) (d \b w + \b h(z, \b z) d\b z) + c(z, \b z)^2 dz d \b z\big)~.}
Here~$w, z$ are holomorphic coordinates and~$K = \d_w$. This metric describes a two-torus fibered over a Riemann surface~$\Sigma$ with metric~$ds^2_\Sigma = \Omega^2 c^2 dz d \b z$. The solution~$(\zeta, \t \zeta)$ exists whenever~$\CM$ has metric~\torusfib\ and admits a nowhere vanishing section~$p$ of its anti-canonical bundle~$\b \CK$. Here~$p$ must be invariant under the Killing vector~$K$, but it need not be anti-holomorphic.
\medskip
\noindent In both cases, the background fields~$b^\mu, M$, and~$\t M$ are not completely determined by the geometry of~$\CM$, i.e.\ by the metric and the complex structure. They also depend on the choice of~$p$, which in turn determines the spinors~$\zeta$ and~$\t \zeta$.

In section~4 we analyze Riemannian four-manifolds~$\CM$ that admit a supercharge~$(\zeta, \t \zeta)$, such that the Killing vector~$K^\mu = \zeta \sigma^\mu \t \zeta$ does not commute with its complex conjugate, $[K, \b K] \neq 0$. These manifolds are locally isometric to a warped product~$S^3\times\R$ with metric
\eqn\warpi{ds^2 = d\tau^2 + r^2(\tau) d \Omega_3~,}
where~$d\Omega_3$ is the round metric on the unit three-sphere. The metric~\warpi\ describes a three-sphere whose radius~$r(\tau)$ varies along~$\R$, and hence the isometry group is~$SU(2) \times SU(2)$. We can solve for~$(\zeta, \t \zeta)$ on any such manifold.

As before, the background fields~$b^\mu, M$, and~$\t M$ are not completely determined by the metric~\warpi. Consequently, they need not be invariant under the full~$SU(2) \times SU(2)$ isometry. If we choose them to respect the full isometry group, there exists a second independent supercharge. Together with~$(\zeta, \t \zeta)$ it generates the super algebra~$OSp(1|2)$. When the domain of~$\tau$ is a compact interval and~$r(\tau)$ vanishes at its endpoints, so that the three-sphere smoothly shrinks to zero size, then the metric~\warpi\ describes a certain squashed four-sphere. In this case the spinors have isolated zeros and~$\CM$ is not a Hermitian manifold.

In section~5 we consider Hermitian manifolds that admit two supercharges of the kind discussed in section~3, focusing on the cases where both supercharges square to zero. (The details of the remaining cases are discussed in appendix~C.) If these two supercharges anti-commute, then~$\CM$ is locally conformal to a Calabi-Yau manifold or to~$H^3 \times \R$, where~$H^3$ is the three-dimensional hyperbolic space of constant negative curvature.\foot{Note, however, that in our conventions the scalar curvature of~$H^3$ with radius~$r$ is~$R = {6 \over r^2}$. By contrast, a three-sphere of radius~$r$ has~$R = - {6 \over r^2}$.} If~$\CM$ is compact, we can say more. In this case it must be globally conformal to a flat torus~$T^4$ or to a~$K3$ surface with Ricci-flat K\"ahler metric. If the two supercharges square to zero but do not anti-commute with one another, the Hermitian metric on~$\CM$ is of the form~\torusfib, with~$h$ determined in terms of~$\Omega$ and~$c$.

In section~6 we briefly discuss manifolds admitting four supercharges. Our conventions are summarized in appendix~A. In appendix~B we review some useful facts about the Chern connection on Hermitian manifolds. Appendix~C contains supplementary material related to section~5.

\newsec{General Properties of the Equations}

We will study the equations~\speq\ on a smooth, oriented, connected four-manifold~$\CM$, which carries a Riemannian metric~$g_{\mu\nu}$. Locally, the spinors~$\zeta$ and~$\t \zeta$ transform as~$(\half, 0)$ and~$(0,\half)$ representations of~$SU(2)_+ \times SU(2)_-$ frame rotations. Globally, they are sections of the left-handed and right-handed spin bundles~$S_+$ and~$S_-$. In order for these bundles to be well defined, $\CM$ must also be endowed with a spin structure.

In order to analyze the restrictions on the background fields due to the presence of one or several solutions of~\speq, we will consider various spinor bilinears. After introducing these bilinears at a point and listing some of their properties that follow from Fierz identities alone, we will use the equations~\speq\ to study their derivatives. In particular, we will investigate the conditions under which the spinors are nowhere vanishing on~$\CM$.

\subsec{Spinor Bilinears}

In this subsection we closely follow the analogous discussion in~\DumitrescuHE. We will work at a point on~$\CM$ and assume that both~$\zeta$ and~$\t \zeta$ are non-zero. The norms~$|\zeta|^2$ and~$|\t \zeta|^2$ are positive scalars. We can thus define two almost complex structures
\eqn\acs{J_{\mu\nu} = {2 i \over |\zeta|^2} \zeta^\dagger \sigma_{\mu\nu} \zeta~, \qquad \t J_{\mu\nu} = {2 i \over |\t \zeta|^2} \t \zeta^\dagger \t \sigma_{\mu\nu} \t \zeta~,}
which satisfy
\eqn\jjid{{J^\mu}_\nu {J^\nu}_\rho = {\, \t J^\mu}_\nu {\, \t J^\nu}_\rho = - {\delta^\mu}_\rho~.}
Both almost complex structures are compatible with the metric, i.e.~$g_{\mu\nu} {J^\mu}_\alpha {J^\nu}_\beta  = g_{\alpha\beta}$ and similarly for~${\, \t J^\mu}_\nu$, and hence they define almost Hermitian structures. Note that~$J_{\mu\nu}$ is self-dual, while~${\t J}_{\mu\nu}$ is anti-self-dual. A vector~$U^\mu$ is holomorphic with respect to~${J^\mu}_\nu$, i.e.\ ${J^\mu}_\nu U^\nu = i U^\mu$, if and only if~$U^\mu \t \sigma_\mu \zeta = 0$. Similarly, $U^\mu$ is holomorphic with respect to~${\, \t J^\mu}_\nu$ if and only if~$U^\mu \sigma_\mu \t \zeta = 0$. (See for instance~\LawsonYR.)

We will also need the two-forms
\eqn\pdef{P_{\mu\nu} = \zeta \sigma_{\mu\nu} \zeta~, \qquad \t P_{\mu\nu} = \t \zeta \, \t \sigma_{\mu\nu} \t \zeta~,}
which are self-dual and anti-self-dual respectively. Since
\eqn\pptholc{{J_\mu}^\nu P_{\nu\rho} = i P_{\mu\rho}~,\qquad {\, \t J_\mu}^\nu \t P_{\nu\rho} = i \t P_{\mu\rho}~,}
it follows that~$P_{\mu\nu}$ is anti-holomorphic with respect to~${J^\mu}_\nu$ and~$\t P_{\mu\nu}$ is anti-holomorphic with respect to~${\, \t J^\mu}_\nu$.

Finally, we consider the complex vectors
\eqn\vecs{K^\mu = \zeta \sigma^\mu \t \zeta~, \qquad X^\mu = \zeta \sigma^\mu \t \zeta^\dagger~,}
which satisfy
\eqn\holc{\eqalign{& {J^\mu}_\nu K^\nu = {\, \t J^\mu }_\nu K^\nu = i K^\mu~, \cr
&  {J^\mu}_\nu X^\nu = -{\, \t J^\mu}_\nu X^\nu = i X^\mu~.}}
Therefore, $K^\mu$ is holomorphic with respect to both~${J^\mu}_\nu$ and~${\, \t J^\mu}_\nu$, but~$X^\mu$ is holomorphic with respect to~${J^\mu}_\nu$ and anti-holomorphic with respect to~${\, \t J^\mu}_\nu$.

The only non-vanishing inner products between~$K^\mu, X^\mu$ and their complex conjugates~$\b K^\mu, \b X^\mu$ are given by
\eqn\norms{\b K^\mu K_\mu = \b X^\mu X_\mu = 2 |\zeta|^2 |\t \zeta|^2~.}
Therefore, these four complex vectors form a complete basis. Projecting onto this basis, we obtain the following useful formulas:
\eqn\ids{\eqalign{& g_{\mu\nu} = {1 \over 2 |\zeta|^2 |\t \zeta|^2} \left(K_\mu \b K_\nu + K_\nu \b K_\mu + X_\mu \b X_\nu + X_\nu \b X_\mu\right)~,\cr
& J_{\mu\nu} = {i \over 2 |\zeta|^2 |\t \zeta|^2} \left(K_\mu \b K_\nu - K_\nu \b K_\mu + X_\mu \b X_\nu - X_\nu \b X_\mu\right)~,\cr
& \t J_{\mu\nu} = {i \over 2 |\zeta|^2 |\t \zeta|^2} \left(K_\mu \b K_\nu - K_\nu \b K_\mu - X_\mu \b X_\nu + X_\nu \b X_\mu\right)~,\cr
& P_{\mu\nu} = {1 \over 2 |\t \zeta|^2} \left(K_\mu X_\nu - K_\nu X_\mu\right)~,\cr
& \t P_{\mu\nu} = -{1 \over 2|\zeta|^2} \left(K_\mu \b X_\nu - K_\nu \b X_\mu\right)~.}}
Since~$J_{\mu\nu}$ is self-dual and~$\t J_{\mu\nu}$ is anti-self-dual, we can also write
\eqn\sdeq{\eqalign{& J_{\mu\nu} = I_{\mu\nu} + \half \ep_{\mu\nu\rho\lambda} I^{\rho\lambda}~,\qquad \t J_{\mu\nu} = I_{\mu\nu} - \half \ep_{\mu\nu\rho\lambda} I^{\rho\lambda}~,\cr
&  I_{\mu\nu} = {i \over \b K^\lambda K_\lambda} \left(K_\mu \b K_\nu - K_\nu \b K_\mu\right)~,}}
so that~$J_{\mu\nu}$ and~${\t J}_{\mu\nu}$ are completely determined in terms of~$K_\mu$ alone. Similarly, the vector~$X^\mu$ is completely determined in terms of~$K^\mu$ and~$P_{\mu\nu}$.

\subsec{Global Properties}

We will now assume that the pair~$(\zeta, \t \zeta)$ satisfies~\speq,
\eqn\speqbis{\eqalign{& \grad_\mu \zeta = {i \over 6} M \sigma_\mu \t \zeta + {i \over 3} b_\mu \zeta + {i \over 3} b^\nu \sigma_{\mu\nu} \zeta~,\cr
& \grad_\mu \t \zeta = {i \over 6} \t M \t \sigma_\mu \zeta - {i \over 3} b_\mu \t \zeta - {i \over 3} b^\nu \t \sigma_{\mu\nu} \t \zeta~.}}
These equations are linear, homogeneous and first order, with smooth coefficients. It follows that the solution~$(\zeta, \t \zeta)$ is determined by its value at a point, and hence the solutions have the structure of a complex vector space of dimension~$\leq 4$. In particular, if~$\zeta$ and~$\t \zeta$ vanish at the same point, then both vanish everywhere on~$\CM$ and the solution is trivial. However, they may vanish one at a time.
As we saw in the previous subsection, it is possible to construct an almost complex structure on~$\CM$ whenever one of the spinors is nowhere vanishing. We will now investigate when this is the case.

Consider the complex vector~$K^\mu = \zeta \sigma^\mu \t \zeta$. The fact that~$\zeta$ and~$\t \zeta$ satisfy~\speqbis\ allows us to compute the derivative of~$K^\mu$ and show that it is a Killing vector,
\eqn\kkv{\grad_\mu K_\nu + \grad_\nu K_\mu = 0~.}
Since the metric is real, it follows that the complex conjugate~$\b K^\mu$ is also a Killing vector, and hence their commutator gives rise to a third, real Killing vector~$L^\mu$,
\eqn\kkbcom{[K, \b K] = 4 i L~.}
Using~\speqbis\ to compute the left-hand side, we can express~$L^\mu$ as follows:
\eqn\lfml{L^\mu = \lambda X^\mu + \b \lambda \b X^\mu~, \qquad \lambda = {1 \over 12} \big( \t M |\zeta|^2-  \b M | \t \zeta|^2 + (b_\nu - \b b_\nu)\b X^\nu\big)~.}
We distinguish two qualitatively different cases, depending on whether~$K$ and~$\b K$ commute:

\medskip

\item{1.)} If~$K$ and~$\b K$ do not commute, then~$L$ is a non-trivial Killing vector. This case will be discussed in section~4. As we will see, the presence of the three Killing vectors~$K, \b K$, and~$L$ implies that~$\CM$ is locally isometric to warped~$S^3 \times \R$ with metric
\eqn\warpmet{ds^2 = d\tau^2 + r^2(\tau) d\Omega_3~.}
Whenever the three-sphere shrinks to zero size, one of the spinors vanishes.

\item{2.)} If~$K$ and~$\b K$ commute, then $L = 0$ everywhere on~$\CM$. In this case~$\zeta$ either vanishes identically or nowhere on~$\CM$, and similarly for~$\t \zeta$.

\medskip
\noindent To prove~$2.)$ we assume that~$\zeta(x) = 0$ for some point~$x \in \CM$. We will first show that~$\grad_\mu \zeta(x) = 0$ as well. Unless the solution is trivial, the fact that~$\zeta(x) = 0$ implies that~$\t \zeta(x) \neq 0$, and hence that~$\t \zeta \neq 0$ in sufficiently small neighborhoods of~$x$. If~$\zeta$ vanishes identically in such neighborhoods, then~$\grad_\mu \zeta(x) = 0$. Otherwise, they contain points at which~$\zeta \neq 0$. At such points, both~$\zeta \neq 0$ and~$\t \zeta \neq 0$, so that~$X^\mu \neq 0$ as well. Since~$L^\mu$ is assumed to vanish identically, it follows from~\lfml\ that~$\lambda = 0$ at these points, and hence that~$\lambda(x) = 0$. From the explicit form of~$\lambda$ in~\lfml, we see that~$M(x) = 0$. Using the first equation in~\speqbis, we conclude that~$\grad_\mu \zeta (x) = 0$.

We can now show that~$\zeta$ vanishes everywhere on~$\CM$. Since~$K^\mu = \zeta \sigma^\mu \t \zeta$, it follows from~$\zeta(x) = 0$ and~$\grad_\mu \zeta(x) = 0$ that~$K_\mu (x) = 0$ and~$\grad_\mu K_\nu(x) = 0$. But~$K^\mu$ is a Killing vector, and hence it must vanish everywhere on~$\CM$.\foot{Every Killing vector~$K^\mu$ satisfies the identity
\eqn\kvid{\grad_\mu \grad_\nu K_\rho = {R_{\nu\rho\mu}}^\lambda K_\lambda~,}
so that~$K_\mu$ is completely determined by specifying~$K_\mu$ and~$\grad_\mu K_\nu$ at a point.} We conclude that the zero set of~$\zeta$ and the zero set of~$\t \zeta$ cover all of~$\CM$. Moreover, they are disjoint, unless the solution is trivial. Since~$\CM$ is connected and the zero set of~$\zeta$ is non-empty, this set must coincide with all of~$\CM$. Therefore~$\zeta$ vanishes everywhere on~$\CM$. Similarly, if~$\t \zeta$ vanishes at a point, it must vanish everywhere on~$\CM$.

Whenever~$\zeta$ is nowhere vanishing, we can use~\acs\ to define an almost complex structure~${J^\mu}_\nu$ that is compatible with the metric. We will now show that~${J^\mu}_\nu$ is integrable, so that~$\CM$ is a Hermitian manifold. It is straightforward (but tedious) to compute~$\grad_\mu {J^\nu}_\rho$ using~\speqbis, and to show that the Nijenhuis tensor~${N^\mu}_{\nu\rho}$ of~${J^\mu}_\nu$ vanishes,
\eqn\nt{{N^\mu}_{\nu\rho} =  {J^\lambda}_\nu \grad_\lambda {J^\mu}_\rho - {J^\lambda}_\rho \grad_\lambda {J^\mu}_\nu -{J^\mu}_\lambda \grad_\nu {J^\lambda}_\rho + {J^\mu}_\lambda \grad_\rho {J^\lambda}_\nu = 0~.}
Therefore~${J^\mu}_\nu$ is an integrable complex structure. Similarly, ${\, \t J^\mu}_\nu$ is an integrable complex structure whenever~$\t \zeta$ is nowhere vanishing.

Alternatively, we can prove that~${J^\mu}_\nu$ is integrable by showing that the commutator of two holomorphic vectors is also holomorphic. Recall from the previous subsection that a vector~$U^\mu$ is holomorphic with respect to~${J^\mu}_\nu$ if and only if~$U^\mu \t \sigma_\mu \zeta = 0$. By differentiating this formula, contracting with another holomorphic vector~$V^\mu$, and antisymmetrizing, we find that the commutator~$[U,V]$ is holomorphic if and only if
\eqn\uvcom{U^{[\mu} V^{\nu]} \t \sigma_\mu \grad_\nu \zeta = 0~.}
Using~\speqbis\ and the fact that~$U^\mu V_\mu = 0$, because~$U^\mu$ and~$V^\mu$ are holomorphic, we can rewrite~\uvcom\ as
\eqn\jint{M U^\mu V^\nu \t \sigma_\mu \sigma_\nu \t \zeta = 0~.}
It suffices to check this condition at points where~$\t \zeta \neq 0$, so that we can take~$U^\mu = X^\mu$ and~$V^\mu = K^\mu$, which form a basis for holomorphic vectors at such points. Since~$K^\mu \sigma_\mu \t \zeta = 0$ by a Fierz identity, we conclude that~${J^\mu}_\nu$ is integrable. A similar argument shows that~${\, \t J^\mu}_\nu$ is also integrable, whenever it exists.

When~$\zeta$ is nowhere vanishing, the complex structure~${J^\mu}_\nu$ splits the bundle of two-forms into~$\Lambda^{2,0} \oplus \Lambda^{1,1} \oplus \Lambda^{0,2}$. Here~$\CK = \Lambda^{2,0}$ is the canonical line bundle of~$(2,0)$ forms. Its complex conjugate~$\b \CK = \Lambda^{0,2}$ is the anti-canonical line bundle of~$(0,2)$ forms. It follows from~\pptholc\ that the two-form~$P_{\mu\nu}$ defined in~\pdef\ is a nowhere vanishing section of~$\b \CK$, and hence that the bundle~$\b \CK$ is trivial. Similarly, a nowhere vanishing~$\t \zeta$ gives rise to a nowhere vanishing section~$\t P_{\mu\nu}$ of the anti-canonical bundle corresponding to~${\, \t J^\mu}_\nu$.

\newsec{Manifolds Admitting One Supercharge}

In this section we will analyze manifolds~$\CM$ that admit one solution~$(\zeta, \t \zeta)$ of~\speq. Here we limit ourselves to the case where the Killing vector~$K^\mu = \zeta \sigma^\mu \t \zeta$ commutes with its complex conjugate~$\b K^\mu$. (Solutions with~$[K, \b K] \neq 0$ will be discussed in section~4.) As we saw in the previous section, this implies that~$\CM$ is a Hermitian manifold. The solutions fall into two classes, which will be discussed in turn:
\medskip

\item{1.)} Solutions of the form~$(\zeta, 0)$ with~$\zeta$ nowhere vanishing and~$\t \zeta$ identically zero, and similarly with the roles of~$\zeta$ and~$\t \zeta$ interchanged. In this case~$K^\mu$ vanishes identically.

\item{2.)} Solutions of the form~$(\zeta, \t \zeta)$ with both~$\zeta$ and~$\t \zeta$ nowhere vanishing. In this case, the Hermitian metric on~$\CM$ is constrained by the presence of the nowhere vanishing Killing vector~$K^\mu$.

\subsec{One Supercharge of the Form~$(\zeta,0)$}

If~$\t \zeta$ vanishes identically, then~\speq\ reduces to
\eqn\bonly{\grad_\mu \zeta = {i \over 3} b_\mu \zeta + {i \over 3} b^\nu \sigma_{\mu\nu} \zeta~,}
with~$\t M=0$ and~$M$ arbitrary. Such a solution corresponds to a supercharge~$Q=(\zeta, 0)$ that squares to zero, $\delta_Q^2 = 0$. Any non-trivial solution~$\zeta$ of~\bonly\ is nowhere vanishing. As in section~2, such a solution gives rise to an integrable Hermitian structure~${J^\mu}_\nu$ and a nowhere vanishing section~$P_{\mu\nu}$ of the corresponding anti-canonical bundle~$\b \CK$ of~$(0,2)$ forms.

We can use the complex structure to introduce holomorphic coordinates~$z^i \, (i = 1,2)$. (Holomorphic and anti-holomorphic indices will be denoted by unbarred and barred indices respectively, e.g.\ $i$ and~$\b i$.) In these coordinates, the complex structure takes the form
\eqn\jcoord{{J^i}_j = i {\delta^i}_j~, \qquad {J^{\b i}}_{\b j} = - i {\delta^{\b i}}_{\b j}~.}
The only non-vanishing components of the metric are~$g_{i \b j} = g_{\b j i}$. Since~$P_{\mu\nu}$ is a~$(0,2)$ form, it only carries barred indices. In a given coordinate patch, we define~$p = P_{\b 1\b2}$. Under a holomorphic coordinate change,~$p$ transforms as follows:
\eqn\ptransf{z'^i = z'^i(z)~, \qquad p'(z') = p(z) \, \det \bigg( \b {\d z'^i \over \d z^j}\bigg)^{-1} ~.}

In general, the complex structure~${J^\mu}_\nu$ is not covariantly constant with respect to the Levi-Civita connection~$\grad_\mu$. (This is only the case if the manifold is K\"ahler.) Instead, we will work with the Chern connection~$\grad_\mu^c$, which has the property that~$\grad_\mu^c g_{\nu\rho} = 0$ and~$\grad_\mu^c {J^\nu}_\rho = 0$. (Some useful properties of the Chern connection are summarized in appendix~B.) The Chern connection acts on sections of the anti-canonical bundle in a simple way,
\eqn\chac{\grad_i^c p = \d_i p~, \qquad \grad_{\b i}^c p = \d_{\b i} p - {p \over 2} \d_{\b i} \log g~,}
where~$g = \det (g_{\mu\nu})$.

We will first determine the form of~$b_\mu$ in the presence of a non-trivial solution~$\zeta$.  Using~\bonly\ we compute
\eqn\dj{\grad_\mu {J^\mu}_\nu = {1 \over 3} (b_\nu + \b b_\nu) - {i\over 3}  (b_\mu - \b b_\mu) {J^\mu}_\nu~,}
so that~$b_\mu$ can be expressed as
\eqn\bsolve{b_\mu = \half (2 g_{\mu\nu} + i J_{\mu\nu}) \grad_\rho J^{\rho\nu} + b^c_\mu~, \qquad {J_\mu}^\nu b^c_\nu = i b^c_\mu~.}
We can now rewrite~\bonly\ in terms of~$b_\mu^c$ and the Chern connection,
\eqn\bonlyii{(\grad_\mu^c - {i \over 2} b^c_\mu)\zeta = 0~.}
Using the fact that~$p = \zeta \sigma_{\b 1 \b 2} \zeta$, this implies that
\eqn\pdiv{(\grad^c_\mu - i b^c_\mu) p = 0~,}
and hence we can solve for~$b^c_\mu$ in terms of~$p$,
\eqn\Bsolve{b^c_\mu = - i \grad_\mu^c \log p~.}
Since~\bsolve\ requires that~$b^c_i = 0$, we see from~\chac\ that~$p$ must be anti-holomorphic,
\eqn\pah{\d_i p = 0~,}
while~$b^c_{\b i}$ is given by
\eqn\conc{b^c_{\b i} = - i \d_{\b i} \log (p g^{-\half})~.}
Under a holomorphic coordinate change~\ptransf, the product~$p g^{-\half}$ is multiplied by a holomorphic function, and hence~$b^c_\mu$ transforms as a well-defined one-form.

In summary, we can solve for~$b_\mu$ on any Hermitian manifold that admits a nowhere vanishing, anti-holomorphic section~$p$ of the anti-canonical bundle~$\b \CK$. In this case~$b_\mu$ is determined by the complex structure~${J^\mu}_\nu$, the Hermitian metric~$g_{\mu\nu}$, and~$p$ according to~\bsolve\ and~\Bsolve.

Conversely, given this data, we can find a solution~$\zeta$ of~\bonly. We introduce a holomorphic frame,
\eqn\frameh{
{1 \over \sqrt 2} e^1=\sqrt{g_{1\b1}} \, dz^1+{g_{2\b 1}\over \sqrt{g_{1\b 1}}} \, dz^2~, \qquad {1 \over \sqrt 2} e^2={g^{1\over 4}\over \sqrt{g_{1\b 1}}} \, dz^2~,}
such that
\eqn\vielbein{ds^2=e^1 e^{\b 1}+e^2 e^{\b 2}~.}
In this frame, the solution of~\bonly\ with~$b_\mu$ given by~\bsolve\ and~\Bsolve\ takes the form
\eqn\zetasol{\zeta_\alpha ={\sqrt s\over 2}\pmatrix{0 \cr 1}~, \qquad s = p g^{-{1 \over 4}}~.}
Using~\ptransf, we see that~$s$ transforms by a phase under holomorphic coordinate changes.

The solution~\zetasol\ is valid locally, in a given coordinate patch. Due to the presence of~$\sqrt p$ in~\zetasol, we have to choose a branch of the square-root in every patch. The requirement that~$\zeta$ be a globally well-defined, smooth section of the spin bundle~$S_+$ then fixes the spin structure on~$\CM$ in terms of the section~$p$. This reflects the fact that square-roots of the anti-canonical bundle~$\b \CK$ correspond to spin structures on~$\CM$. (See for instance~\LawsonYR.)

As a simple example, consider~$\R^3 \times S^1$, obtained from flat~$\C^2$ with holomorphic coordinates~$w, z$ by identifying~$z \sim z + 2\pi i$. This space admits two inequivalent spin structures, corresponding to periodic or anti-periodic boundary conditions for spinors around the compact~$S^1$. If we pick the constant section~$p(\b w, \b z)=1$, it follows from~\zetasol\ that~$\zeta$ is locally constant. In order to ensure that~$\zeta$ is smooth, we must choose the periodic spin structure. We see from~\bsolve\ and~\Bsolve\ that~$b_\mu$ vanishes, and hence this case corresponds to conventional flat-space supersymmetry with one dimension compactified on a circle. As is well-known, this requires periodic boundary conditions for the spinors. On the other hand, choosing~$p(\b w, \b z)=e^{\b z}$ results is a~$\zeta$ that accumulates a sign as it winds around the~$S^1$, and hence we must pick the anti-periodic spin structure. In this case the non-trivial~$\b z$-dependence of~$p(\b w, \b z)$ implies a non-zero value for the background field~$b_\mu$.

If~$\CM$ is compact, the requirement that it admit a nowhere vanishing, anti-holomorphic section of the anti-canonical bundle~$\b \CK$ is very restrictive. In the Enriques-Kodaira classification of compact complex surfaces, only tori, $K3$ surfaces, and primary Kodaira surfaces have this property~\Barth. The Hopf surface~$S^3 \times S^1$ does not admit such a section, but its non-compact version~$S^3 \times \R$ does.

It is straightforward to repeat the preceding analysis for solutions of the form~$(0, \t \zeta)$ with~$\t \zeta$ nowhere vanishing.

\subsec{One Supercharge of the Form~$(\zeta, \t \zeta)$ with~$[K, \b K] = 0$}

We now consider manifolds~$\CM$ that admit a non-trivial solution~$(\zeta, \t \zeta)$ of~\speq,
\eqn\speqtris{\eqalign{& \grad_\mu \zeta = {i \over 6} M \sigma_\mu \t \zeta + {i \over 3} b_\mu \zeta + {i \over 3} b^\nu \sigma_{\mu\nu} \zeta~,\cr
& \grad_\mu \t \zeta = {i \over 6} \t M \t \sigma_\mu \zeta - {i \over 3} b_\mu \t \zeta - {i \over 3} b^\nu \t \sigma_{\mu\nu} \t \zeta~.}}
Here we restrict ourselves to solutions for which the Killing vector~$K^\mu= \zeta \sigma^\mu \t \zeta $ is non-zero and commutes with its complex conjugate, $[K, \b K] = 0$. As we showed in the previous section, this implies that both~$\zeta$ and~$\t \zeta$ are nowhere vanishing, and hence the same is true for~$K^\mu$. The two spinors determine two complex structures~${J^\mu}_\nu$ and~${\, \t J^\mu}_\nu$ according to~\acs, both of which are compatible with the metric. Alternatively, we see from~\sdeq\ that~${J^\mu}_\nu$ and~${\, \t J^\mu}_\nu$ are completely determined in terms of~$K^\mu$. Moreover, it follows from~\holc\ that~$K^\mu$ is holomorphic with respect to both complex structures.

The spinors~$\zeta$ and~$\t \zeta$ give rise to nowhere vanishing two-forms~$P_{\mu\nu}$ and~$\t P_{\mu\nu}$, which are sections of the anti-canonical bundles corresponding to~${J^\mu}_\nu$ and~${\, \t J^\mu}_\nu$ respectively. However, it follows from~\ids\ that~$K^\mu$ and~$P_{\mu\nu}$ completely determine the vector~$X^\mu = \zeta \sigma^\mu \t \zeta^\dagger$, and hence also~$\t P_{\mu\nu}$. Therefore, all spinor bilinears are completely determined by specifying~$K^\mu$ and~$P_{\mu\nu}$. Finally, it follows directly from~\speqtris\ that~$P_{\mu\nu}$ and~$\t P_{\mu\nu}$ are invariant along~$K^\mu$,
\eqn\lkp{\CL_K P_{\mu\nu} = 0~, \qquad \CL_K \t P_{\mu\nu} = 0~.}

Using the complex structure~${J^\mu}_\nu$, we introduce holomorphic coordinates~$w, z$ such that~$K = \d_w$. This restricts the form of the metric,
\eqn\ellmet{ds^2 =\Omega(z,\bar z)^2\left( (dw +h(z,\bar z) dz)(d\bar w +\b h(z,\bar z) d\bar z)+ c(z,\bar z)^2 dz d\bar z\right)~,}
which describes a two-torus fibered over a Riemann surface~$\Sigma$ with metric~$d s^2_\Sigma = \Omega^2 c^2 dz d\b z$. Note that the conformal factor~$\Omega^2$ is determined by the norm of~$K^\mu$, which in turn depends on the norms of~$\zeta$ and~$\t \zeta$,
\eqn\knorm{\Omega^2 = 2 \b K^\mu K_\mu = 4 |\zeta|^2 |\t \zeta|^2~.}

We will now determine the form of the background fields~$b_\mu, M$, and~$\t M$ in the presence of a solution~$(\zeta, \t \zeta)$. In order to constrain~$b_\mu$, we compute using~\speqtris,
\eqn\djjt{\grad_\mu {J^\mu}_\nu = {1 \over 3} (b_\nu + \b b_\nu) -{i \over 3} (b_\mu - \b b_\mu) {J^\mu}_\nu  + {1 \over 3|\zeta|^2} (\b M X_\nu + M \b X_\nu)~.}
This restricts~$b_\mu$ to be of the form
\eqn\bsolveii{b_\mu = {3 \over 2} \grad_\nu {J^\nu}_\mu - {1 \over 2 |\zeta|^2 |\t \zeta|^2} (\t M |\zeta|^2 X_\mu + M |\t \zeta|^2 \b X_\mu) + B_\mu~, \qquad {J^\mu}_\nu B^\nu = iB^\mu~.}
As in the previous subsection, we substitute this into~\speqtris\ and find that the equation satisfied by~$\zeta$ reduces to~\bonlyii\ with~$b_\mu^c$ given by
\eqn\bc{b_\mu^c = b_\mu - \half (2 g_{\mu\nu} + i J_{\mu\nu} ) \grad_\rho J^{\rho\nu}~.}
However, $b_\mu^c$ is no longer required to be holomorphic, due to the presence of the terms proportional to~$M$ and~$\t M$ in~\bsolveii. It follows that~$p=P_{\b w \b z}$ satisfies~\pdiv, and hence~$b_\mu^c$ is given by~\Bsolve. In~$w, z$ coordinates,
\eqn\bccomp{b^c_w = 0~, \qquad b^c_z = - i \d_z p~, \qquad b^c_{\b i} = - i \d_{\b i} \log (p g^{- \half})~,}
where first equation follows from~\lkp\ and the fact that~$K = \d_w$.

In order to determine~$M$ and~$\t M$, we use~\speqtris\ to compute
\eqn\gradp{\grad^\mu P_{\mu\nu} = { i \over 2} M K_\nu~, \qquad \grad^\mu \t P_{\mu\nu} = {i \over 2} \t M K_\nu~,}
so that
\eqn\msolve{M = - {2 i \over \b K^\rho K_\rho} \b K^\nu \grad^\mu P_{\mu\nu}~, \qquad \t M = - {2 i \over \b K^\rho K_\rho} \b K^\nu \grad^\mu \t P_{\mu\nu}~.}
Using the expressions~\ids\ for~$P_{\mu\nu}$ and~$\t P_{\mu\nu}$ in terms of~$K^\mu$ and~$X^\mu$, and the fact that the Killing vector~$K^\mu$ commutes with its complex conjugate, we can rewrite~\msolve\ as follows:
\eqn\mmt{M = - i \grad^\mu \left({\b X_\mu \over |\zeta|^2}\right)~, \qquad \t M = i \grad^\mu \left({X_\mu \over |\t \zeta|^2}\right)~.}
This emphasizes that~$M$ and~$\t M$ are naturally viewed as dual four-form field strengths. Below, we will also need explicit formulas for~$M$ and~$\t M$ in~$w, z$ coordinates,
\eqn\mmbcoord{M = {2 i p \over \Omega^4 c^2}\,  \d_z \log p~, \qquad \t M = - { 2i \Omega^2 \over p} \left(\d_{\b z} \log\bigg({\Omega^6 c^2 \over p }\bigg) + \b h \, \d_{\b w} \log p \right)~.}
Here~$\Omega(z, \b z), h(z, \b z)$, and~$c(z, \b z)$ are the functions appearing in the metric~\ellmet\ and~\lkp\ implies that~$p(\b w, z, \b z)$ does not depend on~$w$. The formulas~\mmbcoord\ explicitly show that~$M$ and~$\t M$ are completely determined in terms of the Hermitian metric and~$p$.

In summary, we have solved for the background fields~$b_\mu, M$, and~$\t M$ on any Hermitian manifold with metric~\ellmet\ and a nowhere vanishing section~$p$ of the anti-canonical bundle~$\b \CK$ that is invariant under the holomorphic Killing vector~$K = \d_w$. The background fields are then determined in terms of the complex structure, the Hermitian metric, and~$p$ according to~\bc, \bccomp, and~\mmbcoord. Even though we have explicitly worked in terms of the complex structure~${J^\mu}_\nu$ and the section~$P_{\mu\nu}$ of the corresponding anti-canonical bundle, we could have also phrased the entire discussion in terms of~${\, \t J^\mu}_\nu$ and~$\t P_{\mu\nu}$ to obtain an equivalent set of formulas for the background fields.

Conversely, we can find a solution~$(\zeta, \t \zeta)$ of~\speqtris\ on any manifold that admits a nowhere vanishing complex Killing vector~$K^\mu$, which squares to zero, $K^\mu K_\mu = 0$, and commutes with its complex conjugate, $[K, \b K] = 0$. As we saw above, the presence of~$K^\mu$ allows us to define a complex structure~${J^\mu}_\nu$ and restricts the metric to be of the form~\ellmet. If we are given a nowhere vanishing section~$p$ of the corresponding anti-canonical bundle~$\b \CK$ that satisfies~$\CL_K p = 0$, we can explicitly solve for~$\zeta$ and~$\t \zeta$. Introducing a frame adapted to the Hermitian metric~\ellmet\ as in~\frameh,
\eqn\ellframe{e^1 = \Omega (d w + h dz)~, \qquad e^2 = \Omega c dz~,}
we find that~$\zeta$ and~$\t \zeta$ are given by
\eqn\zztell{\zeta_\alpha = {\sqrt s \over 2} \pmatrix{0 \cr 1}~, \qquad \t \zeta^\alphadot = {\Omega \over \sqrt s} \pmatrix{0 \cr 1}~, \qquad s = p g^{- {1 \over 4}}~.}

Finally, we would like to comment on the supersymmetry algebra generated by the supercharge~$Q = (\zeta, \t \zeta)$. It follows from~\salg\ that~$Q$ squares to the complex Killing vector~$K$, which leaves the background fields~$b^\mu, M$, and~$\t M$ invariant. However, the complex conjugate vector~$\b K$ does not appear on the right-hand side of the supersymmetry algebra and it need not be a symmetry of~$b^\mu, M$, and~$\t M$, even though the reality of the metric ensures that it is a Killing vector.

If we would like to add~$\b K$ to the supersymmetry algebra, we must ensure that it leaves~$b^\mu, M$, and~$\t M$ invariant. Given the form of~$b^\mu$ in~\bc\ and~\bccomp, and demanding that~$\CL_{\b K} b^\mu = 0$ fixes the~$\b w$-dependence of~$p$,
\eqn\condp{p(\b w, z, \b z) = e^{\alpha \b w} {\hat p}(z,\bar z)~,\qquad \alpha \in \C~.}
Given this form of~$p$, it follows from~\mmbcoord\ that
\eqn\MinvbK{\CL_{\b K} M= \alpha M~,\qquad \CL_{\b K} \t M= -\alpha  \t M~.}
Demanding that~$M$ and~$\t M$ be invariant under~$\b K$ leads to the following two cases:
\medskip
\item{1.)} If~$\alpha = 0$, the algebra generated by the supercharge~$Q = (\zeta, \t \zeta)$ and the Killing vectors~$K$ and~$\b K$ takes the form
    \eqn\onezeroalg{\eqalign{& \delta_Q^2 = 2 i \delta_K~,\cr
    & [\delta_K, \delta_Q] = [\delta_{\b K}, \delta_Q] = 0~,\cr
    & [\delta_K, \delta_{\b K}] = 0~.}}
    This is the familiar two-dimensional~$(1,0)$ supersymmetry algebra. Here it is geometrically realized on the torus fibers of the metric~\ellmet.

\item{2.)} If~$\alpha \neq 0$, it follows from~\MinvbK\ that~$M = \t M = 0$. The solution~$(\zeta, \t \zeta)$ then splits into two independent supercharges~$Q = (\zeta, 0)$ and~$\t Q = (0, \t \zeta)$.  This case will be discussed in section~5.

\newsec{Solutions with~$[K, \b K] \neq 0$: One or Two Supercharges on Warped~$S^3 \times \R$}

In this section we analyze Riemannian four-manifolds~$\CM$ that admit a solution~$(\zeta, \t \zeta)$ of~\speq, such that the Killing vector~$K^\mu=\zeta \sigma^\mu \tilde \zeta$ does not commute with its complex conjugate, $[K, \b K] \neq 0$. We will show that~$\CM$ is locally isometric to warped~$S^3 \times \R$ with metric
\eqn\metcyl{ds^2=d\tau^2+r(\tau)^2 d\Omega_3~,}
where~$d\Omega_3$ is the round metric on the unit three-sphere. This metric has~$SU(2)_\ell \times SU(2)_r$ isometry and describes a round three-sphere whose radius~$r(\tau)$ varies along~$\R$. This three-sphere can smoothly shrink to zero size at up to two values of~$\tau$. In this case, the metric~\metcyl\ describes a certain squashed four-sphere, which is not a Hermitian manifold. As we will see below, the spinors~$\zeta$ and~$\t \zeta$ have isolated zeros on such manifolds, and hence the almost complex structures defined in~\acs\ do not exist everywhere on~$\CM$.

For any metric of the form~\metcyl, we will determine the background fields~$b^\mu, M, \t M$ and solve for~$(\zeta, \t \zeta)$. As in the previous section, the background fields need not be invariant under the full~$SU(2)_\ell \times SU(2)_r$ isometry. If we choose them to respect the full isometry group, there exists a second independent supercharge.

\subsec{Constraining the Metric}

Here we prove that~$[K, \b K] \neq 0$ implies that the metric must be of the form~\metcyl, following closely the argument in appendix~C of~\DumitrescuHE. By assumption, $K$ and~$\b K$ commute to a third, real Killing vector~$L$,
\eqn\comml{[K,\bar K]=4 i L~.}
First, note that the real vectors~$K + \b K, i(K - \b K)$, and~$L$ are orthogonal. To see this, we differentiate the identity~$K^\mu K_\mu = 0$ along~$\b K$ and use~\comml\ to find
\eqn\proforth{0 = \CL_{\b K} (K^\mu K_\mu) = -8 i L^\mu K_\mu~.}
Therefore~$L^\mu K_\mu = 0$, and hence~$L^\mu \b K_\mu = 0$ as well.

If the three real, orthogonal Killing vectors~$K + \b K, i(K - \b K)$, and~$L$ form a closed algebra, we can obtain further constraints similar to~\proforth\ and show that the algebra is~$SU(2)$ in its usual compact form. Introducing~$SU(2)$-invariant one-forms~$\t \omega^a \, (a = 1, 2, 3)$, we can write the metric as~$ds^2 = d\tau^2 + (r^2)_{ab} (\tau)  \, \t \omega^a  \t \omega^b$. The fact that the~$SU(2)$ generators~$K + \b K$, $i(K - \b K)$, and~$L$ are orthogonal implies that~$(r^2)_{ab}(\tau) = r(\tau)^2 \delta_{ab}$, so that the metric is given by~\metcyl. Therefore, the isometry is enhanced to~$SU(2)_\ell \times SU(2)_r$ and the~$SU(2)$ generated by~$K, \b K$, and~$L$ is identified with either~$SU(2)_\ell$ or~$SU(2)_r$.

If~$K, \b K$, and~$L$ do not form a closed algebra, we obtain a fourth real Killing vector, which together with the other three generates~$SU(2) \times U(1)$. In this case the metric also takes the form~\metcyl, but the extra~$U(1)$ isometry generates translations in~$\tau$, and hence the radius~$r$ of the~$S^3$ must be constant.

\subsec{One Supercharge}

Given the metric~\metcyl, we will now determine the background fields~$b^\mu, M, \t M$ in the presence of a single supercharge~$(\zeta, \t \zeta)$, before solving for~$(\zeta, \t \zeta)$ itself. Here we assume that~$K, \b K$, and~$L$ generate the~$SU(2)_\ell$ factor of the isometry group. (Switching between~$SU(2)_\ell$ and~$SU(2)_r$ reverses the orientation, which leads to several sign changes below.)  Let us denote by~$\ell_a \, (a = 1, 2, 3)$ the vector fields on the unit three-sphere that generate~$SU(2)_\ell$, normalized so that~$[\ell_a, \ell_b] = - 2 \ep_{abc} \ell_c$. Their dual one-forms~$\omega^a \, (a = 1, 2, 3)$ then furnish an~$SU(2)_r$-invariant frame on the unit three-sphere,\foot{The~$SU(2)_r$-invariant one-forms~$\omega^a$ should not be confused with the~$\t \omega^a$ of the previous subsection, which were invariant under~$K, \b K$, and~$L$. Here the~$\t \omega^a$ would be~$SU(2)_\ell$-invariant one-forms.} so that the metric is~$d\Omega_3 = (\omega^1)^2 + (\omega^2)^2 + (\omega^3)^2$ and the volume form is~$\omega^1 \wedge \omega^2 \wedge \omega^3$. This allows us to define an~$SU(2)_r$-invariant frame for the metric~\metcyl,
\eqn\orthoframe{e^a = r(\tau) \omega^a \quad (a = 1, 2, 3)~,\qquad e^4 = d\tau~.}
Below, we will express the spinors~$\zeta$ and~$\t \zeta$ in this frame.

Up to an overall multiplicative constant, which can be absorbed by rescaling the spinors, we are free to choose
\eqn\kl{K = \ell_1 + i \ell_2~, \qquad L = \ell_3~.}
It follows that~$\b K^\mu K_\mu = 2 r(\tau)^2$, so that~$|\zeta|^2 |\t \zeta|^2 = r(\tau)^2$.  We also introduce a fourth real vector~$T^\mu$,
 \eqn\defZZ{T^\mu=-{i\over \b K^\lambda K_\lambda} \ep^{\mu\nu\rho\sigma }L_\nu K_\rho \b K_\sigma = r(\tau)\delta^\mu_\tau~,}
which is orthogonal to~$K, \b K$, and~$L$. We can now use the spinors to define a complex function~$s$ as follows:
\eqn\defsfa{|s| = {|\zeta|^2 \over r(\tau)} = {r(\tau) \over |\t \zeta|^2}~, \qquad X^\mu = \zeta \sigma^\mu \t \zeta^\dagger = {s \over |s|} (L^\mu + i T^\mu)~.}
The fact that~$X^\mu$ only differs from~$L^\mu + i T^\mu$ by a phase follows from the expression~\ids\ for the metric in terms of~$K_\mu$ and~$X_\mu$. As in the previous section, the equations~\speq\ imply that the two-form~$P_{\mu\nu} = \zeta \sigma_{\mu\nu} \zeta$ is invariant along~$K$, so that~$\CL_K P_{\mu\nu} = 0$. Using~\defsfa\ and~\ids, we can express~$P_{\mu\nu}$ in terms of the function~$s$. Imposing~$\CL_K P_{\mu\nu} = 0$, we find that~$s$ must be invariant along~$K$,
\eqn\sinv{K^\mu \d_\mu s = 0~.}
However, it need not be invariant along~$\b K$ or~$L$. We will return to this point below.

We can now determine~$b^\mu, M$, and~$\t M$ in terms of geometric quantities and the function~$s$. As in the previous section, we find that~$M$ and~$\t M$ are given by~\msolve. Using~\defsfa\ and~\ids, the answer can be rewritten as follows:
\eqn\solaust{\eqalign{&M=i \grad^\mu\left({s\over r(\tau)}(L_\mu+i T_\mu)\right) - {2 s \over r(\tau)}~,\cr &\t M=-i \grad^\mu\left({1\over s  r(\tau)}(L_\mu-i T_\mu)\right) + {2 \over s r(\tau)}~.}}
To determine~$b_\mu$, we evaluate the commutator~$[X, \b X]$ in two different ways: once directly using~\speq,
\eqn\comW{[X, \b X]= - {i\over 2}  \left(\b X^\mu (b_\mu+\b b_\mu)\right) X - {2i\over 3} (\b K^\mu \b b_\mu) K - ({\rm c.c.})~,}
and once using~\defsfa,
\eqn\comWb{[X, \b X]= - 2 \left(L^\mu \partial_\mu \log {s\over |s|} \right)L - 2 \left(T^\mu \partial_\mu \log {s\over |s|}\right) T~.}
Comparing~\comW\ and~\comWb, we conclude that
\eqn\condp{K^\mu b_\mu=0~,\quad L^\mu(b_\mu+\b b_\mu)=-2 i L^\mu \partial_\mu  \log{s\over |s|}~,\quad T^\mu (b_\mu+\b b_\mu)=-2 i T^\mu \partial_\mu  \log{s\over |s|}~.}
Similarly, we can use the fact that~$|s|^2= {|\zeta|^2} { |\t \zeta|^{-2}}$  to evaluate
\eqn\simbmb{\partial_\mu \log |s|^2=i (b_\mu-\b b_\mu)+{2\over r(\tau)} \delta_\mu^\tau~.}
Together with~\condp, this allows us to solve for~$b_\mu$,
\eqn\solbau{b_\mu=-i \partial_\mu  \log s + { i\over r(\tau)} \, \delta^{\tau}_\mu~.}
It is clear from~\solaust\ and~\solbau\ that the background fields need not be invariant under~$\b K$ or~$L$, if the function~$s$ is not invariant under these isometries.

Having determined~$b^\mu, M$, and~$\t M$ in terms of the metric and the function~$s$, which satisfies~\sinv, we can express the solution~$(\zeta, \t \zeta)$ itself in terms of this data. In the frame~\orthoframe,
\eqn\zzpell{\zeta_\alpha = \sqrt{ s r(\tau) } \pmatrix{1 \cr 0}~, \qquad \t \zeta^\alphadot = {\sqrt{  r(\tau) }\over \sqrt {s}} \pmatrix{-1 \cr 0}~.}

\subsec{Two Supercharges}

As was already mentioned, the function~$s$ and the background fields~$b^\mu, M, \t M$ must be invariant along~$K$, but they need not be invariant along~$\b K$ or~$L$. If we also choose~$s$ to be invariant under~$\b K$ and~$L$, then it can only depend on~$\tau$. The background fields in~\solaust\ and~\solbau\ are then given by
\eqn\simplebgs{\eqalign{& M = - s(\tau) \left({3 r'(\tau) \over r(\tau)} + {s'(\tau) \over s(\tau)} + {2 \over r(\tau)}\right)~,\cr
& \t M = -{1 \over s(\tau)} \left({3 r'(\tau) \over r(\tau)} - {s'(\tau) \over s(\tau)} - {2 \over r(\tau)}\right)~,\cr
& b_\mu = i \left({1 \over r(\tau)} - {s'(\tau) \over s(\tau)}\right) \delta_\mu^\tau~,}}
and hence they are invariant under the full~$SU(2)_\ell \times SU(2)_r$ isometry group. In this case we can find a second independent supercharge,
\eqn\zzpellb{\eta_\alpha = \sqrt{s(\tau) r(\tau)}  \pmatrix{0 \cr -1}~, \qquad \t \eta^\alphadot = \sqrt{r(\tau) \over s(\tau)} \pmatrix{0 \cr 1}~.}
Denoting the two supercharges by~$Q = (\zeta, \t \zeta)$ and~$Q' = (\eta, \t \eta)$, we can use~\salg\ and~\twosc\ to obtain the following anti-commutators:
\eqn\osponetwo{\delta_Q^2 = 2 i \delta_K~, \qquad \delta_{Q'}^2 = - 2i \delta_{\b K}~, \qquad \{\delta_Q, \delta_{Q'}\} = 4 i \delta_L~.}
Similarly, we find that~$Q$ and~$Q'$ are singlets under~$SU(2)_r$ but transform as a doublet of~$SU(2)_\ell$. Therefore, the supersymmetry algebra is given by~$OSp(1|2) \times SU(2)_r$. Here the bosonic subalgebra~$Sp(2) \subset OSp(1|2)$ is identified with the compact~$SU(2)_\ell$ isometry subgroup.

We would now like to consider the situation when the radius~$r(\tau)$ vanishes in such a way that the~$S^3$ smoothly shrinks to zero size. For concreteness, assume that~$r(\tau) \rightarrow 0$ as~$\tau \rightarrow 0^+$. In order to avoid curvature singularities, we must take
\eqn\conrt{r(\tau) = \tau + \CO(\tau^3)~, \qquad \tau \geq 0~.}
Since we also need to ensure that the background fields~$b^\mu, M, \t M$ are smooth, it follows from~\simplebgs\ that the function~$s(\tau)$ satisfies
\eqn\conds{s(\tau) = s_0 \tau + \CO(\tau^3)~,\qquad s_0 \in \C~.}
We thus conclude from~\zzpell\ and~\zzpellb\ that~$\zeta, \eta$ vanish as~$\tau \rightarrow 0^+$, while~$\t \zeta, \t \eta$ do not. If we instead consider the case when~$r(\tau) \rightarrow 0$ as~$\tau \rightarrow 0^-$, then~$r(\tau) = - \tau + \CO(\tau^3)$ and~$\t \zeta, \t \eta$ vanish but~$\zeta, \eta$ do not. Combining these cases, we can take~$\tau_- \leq \tau \leq \tau_+$, such that~$r(\tau_\pm) = 0$. In this case the metric~\metcyl\ describes a squashed four-sphere with~$SU(2)_\ell \times SU(2)_r$ isometry. Each of the spinors vanishes somewhere and hence we cannot use them to construct a complex structure. This is consistent with the fact that these manifolds are not Hermitian.

\subsec{Special Cases}

\item{1.)} If we choose~$r(\tau) = \tau$ with~$\tau \geq 0$, we obtain flat~$\R^4$ in polar coordinates. However, we still have freedom in choosing the function~$s(\tau)$, as long as it satisfies~\conds. Note that no choice of~$s(\tau)$ will make all background fields in~\simplebgs\ vanish. This is because we assumed that~$[K, \b K] \neq 0$, which cannot arise in ordinary flat-space supersymmetry.

\item{2.)} Choosing~$r(\tau) = r_0 \sin{\tau \over r_0}$ results in a round~$S^4$ of radius~$r_0$, and choosing~$r(\tau) = r_0 \sinh {\tau \over r_0}$ gives~$H^4$ of radius~$r_0$. If we set~$s(\tau) = s_0 \tan {\tau \over 2 r_0} $ or~$s(\tau) = s_0 \tanh {\tau \over 2 r_0}$ respectively, we can find two additional supercharges, or four supercharges in total. These cases will be discussed in section~6.

\item{3.)} If~$r(\tau) = r$ is a constant, we obtain~$S^3 \times \R$ with an~$S^3$ of fixed radius~$r$. If we choose~$s(\tau) = s_0 e^{-{2 \over r} \tau}$, then~$M = \t M = 0$, ~$b_\mu = {3 i \over r}\delta_\mu^\tau~$, and the space admits four supercharges (see section 6). However, the spinors vary exponentially along~$\R$, and hence we cannot compactify the~$\tau$-direction to~$S^1$. If we instead choose~$s = s_0$ to be a constant, the spinors are constant as well, and hence we can compactify to an~$S^1$ of any radius. If we let the~$S^1$ shrink to zero size, we obtain a three-dimensional theory with supersymmetry algebra~$OSp(1|2) \times SU(2)_r$ on a round~$S^3$ of radius~$r$.

\newsec{Manifolds Admitting Two Supercharges}

In this section, we consider manifolds that admit two supercharges. We have already encountered such manifolds in the previous section. Here we will discuss Hermitian manifolds that admit two supercharges of the kind discussed in section~3, i.e.\ supercharges that either square to zero or to a Killing vector that commutes with its complex conjugate. Hence, there are four possible cases:
\medskip
\item{1.)} Two supercharges of the form~$(\zeta, 0)$ and~$(\eta, 0)$. (Similarly, two supercharges~$(0, \t \zeta)$ and~$(0,\t \eta)$.) Both supercharges square to zero and anti-commute with one another.
\item{2.)} Two supercharges of the form~$(\zeta, 0)$ and~$(0, \t \zeta)$. Each supercharge squares to zero, but they anti-commute to a Killing vector~$K^\mu = \zeta \sigma^\mu \t \zeta$, and we assume that~$[K, \b K] = 0$.
\item{3.)} Two supercharges of the form~$(\zeta, \t \zeta)$ and~$(\eta, \t \eta)$, both of which square to a non-trivial Killing vector.
\item{4.)} One supercharge of the form~$(\zeta, \t \zeta)$, which squares to a Killing vector, and one supercharge of the form~$(\eta, 0)$ or~$(0, \t \eta)$, which squares to zero.
\medskip
The first two cases are analyzed below. Case~$3.)$ is discussed in appendix~C, where it is shown to only arise on a warped product~$T^3 \times \R$,
\eqn\warpedtt{ds^2 = d \tau^2 + r(\tau)^2 ds^2_{T^3}~,}
or a warped product~$H^3 \times \R$,
\eqn\warpedht{ds^2 = d\tau^2 + r(\tau)^2 ds^2_{H^3}~.}
Here~$ds^2_{T^3}$ is the usual flat metric on the three-torus~$T^3$ and~$ds^2_{H^3}$ is the constant negative curvature metric on~$H^3$ of unit radius.  These warped~$T^3 \times \R$ and warped~$H^3 \times \R$ metrics are the zero and negative curvature analogues of the warped~$S^3 \times \R$ metric~\metcyl\ discussed in the previous section. Case~$4.)$ is even more restrictive. Using arguments similar to those in appendix~C, it can be shown to only arise on~$H^3 \times \R$ with~$H^3$ of constant radius. As we will see in section~6, this space admits four supercharges.

\subsec{Two Supercharges of the Form~$(\zeta, 0)$ and~$(\eta, 0)$}

We begin by presenting a set of integrability conditions that follow from the equations~\speq. (These will also be used in section~6.) Given any solution~$(\zeta, \t \zeta)$ of~\speq, we can use~$\half R_{\mu\nu\rho\lambda} \sigma^{\rho\lambda} \zeta = [\grad_\mu, \grad_\nu] \zeta$ and~$\half R_{\mu\nu\rho\lambda} \t \sigma^{\rho\lambda} \t \zeta = [\grad_\mu, \grad_\nu] \t \zeta$ to obtain the following relations:
\eqn\intcond{\eqalign{& \half R_{\mu\nu\rho\lambda} \sigma^{\rho\lambda} \zeta = {1 \over 9} (M \t M + b^\rho b_\rho) \sigma_{\mu\nu} \zeta + {i \over 3} (\grad_\mu b_\nu - \grad_\nu b_\mu) \zeta - {1 \over 9} M \ep_{\mu\nu\rho\lambda} b^\rho \sigma^\lambda \t \zeta \cr
& \hskip68pt + {i \over 3} \Big( (\grad_\mu b^\rho - {i \over 3} b_\mu b^\rho) \sigma_{\nu\rho} \zeta + \half (\d_\mu M - {2 i \over 3} M b_\mu) \sigma_\nu \t \zeta\,\Big) - (\mu\leftrightarrow\nu)~,\cr
& \half R_{\mu\nu\rho\lambda} \t \sigma^{\rho\lambda} \t \zeta = {1 \over 9} (M \t M + b^\rho b_\rho) \t \sigma_{\mu\nu} \t \zeta - {i \over 3} (\grad_\mu b_\nu - \grad_\nu b_\mu) \t \zeta - {1 \over 9} \t M \ep_{\mu\nu\rho\lambda} b^\rho \t \sigma^\lambda \zeta \cr
& \hskip68pt - {i \over 3} \Big( (\grad_\mu b^\rho + {i \over 3} b_\mu b^\rho) \t \sigma_{\nu\rho} \t \zeta - \half (\d_\mu \t M + {2 i \over 3} \t M b_\mu) \t \sigma_\nu \zeta\,\Big) - (\mu\leftrightarrow\nu)~.}}
We can now specialize to the case of two independent solutions~$(\zeta, 0)$ and~$(\eta, 0)$. Recall from section~2 that such solutions are only possible if~$\t M = 0$, while~$M$ is unconstrained. Substituting into~\intcond\ and using the fact that~$\zeta$ and~$\eta$ are independent at every point, we arrive at the following integrability conditions:
\medskip
\item{1.)} The Weyl tensor is anti-self-dual, $W_{\mu\nu\rho\lambda} = - \half \ep_{\mu\nu\kappa\sigma} {W^{\kappa\sigma}}_{\rho\lambda}$.
\item{2.)} The one-form~$b_\mu$ is closed, $\d_\mu b_\nu - \d_\nu b_\mu = 0$.
\item{3.)} The Ricci tensor is given by
\eqn\ricciint{R_{\mu\nu} = - {i \over 3} (\grad_\mu b_\nu + \grad_\nu b_\mu + g_{\mu\nu} \grad_\rho b^\rho) - {2 \over 9} (b_\mu b_\nu - g_{\mu\nu} b^\rho b_\rho)~.}
\medskip
\noindent If we instead consider two solutions~$(0, \t \zeta)$ and~$(0, \t \eta)$, the Weyl tensor is self-dual rather than anti-self dual, and the Ricci tensor is given by~\ricciint\ with~$b_\mu \rightarrow - b_\mu$.

We will first analyze these integrability conditions locally. Then the fact that~$b_\mu$ is closed implies that it can be written as
\eqn\blocal{b_\mu = b^{(r)}_\mu + 3 i \d_\mu \phi~,}
where~$b_\mu^{(r)}$ is a closed real one-form and~$\phi$ is a real scalar function. If we use~$\phi$ to perform a local conformal rescaling of the metric,
\eqn\metconf{g_{\mu\nu} = e^{2\phi} g'_{\mu\nu}~,}
we find that the new metric~$g_{\mu\nu}'$ satisfies the same integrability conditions~$1.)-3.)$ as above, but with~$b_\mu$ replaced by~$b^{(r)}_\mu$. Since the metric and~$b_\mu^{(r)}$ are real, it follows from~\ricciint\ that~$b_\mu^{(r)}$ is covariantly constant, so that
\eqn\ric{R'_{\mu\nu} = - {2 \over 9} \big(b^{(r)}_\mu b^{(r)}_\nu - g'_{\mu\nu} g'^{\rho\lambda} b^{(r)}_\rho b^{(r)}_\lambda\big)~,\qquad \grad_\mu b^{(r)}_\nu = 0~.}
The solutions therefore fall into two classes:
\medskip
\item{(i)} If~$b_\mu^{(r)} = 0$ then~$R'_{\mu\nu} = 0$ and the Weyl tensor~$W'_{\mu\nu\rho\lambda}$ is anti-self-dual. In this case the whole Riemann tensor~$R'_{\mu\nu\rho\lambda}$ is anti-self-dual, and hence the holonomy of the metric~$g'_{\mu\nu}$ is contained in~$SU(2)$. Therefore~$\CM$ is locally conformal to a Calabi-Yau manifold.

\item{(ii)} If~$b_\mu^{(r)} \neq 0$ it follows from~\ric\ that~$g_{\mu\nu}'$ is locally isometric to~$H^3 \times \R$ with~$b^{(r)}_\mu$ pointing along~$\R$. Here the radius~$r$ of~$H^3$ is determined by~$g'^{\mu\nu} b^{(r)}_\mu b^{(r)}_\nu = {9 \over r^2}$. Therefore~$\CM$ is locally conformal to~$H^3 \times \R$.
\medskip

If~$\CM$ is compact we can say more. In this case~$\CM$ must be globally conformal to either a flat torus~$T^4$ or to a~$K3$ surface with Ricci-flat K\"ahler metric. This follows directly from the results of section~5 in~\DumitrescuHE, where it was shown that the existence of two solutions~$(\zeta, 0)$ and~$(\eta, 0)$ of~\speq\ on a compact manifold~$\CM$ imply that~$\CM$ admits a hyperhermitian structure.\foot{A hyperhermitian structure consists of three complex structures~$J^{(a)} \; (a = 1, 2, 3)$, which are compatible with the same Riemannian metric and satisfy the quaternion algebra,
\eqn\quat{\{J^{(a)}, J^{(b)}\} = - 2 \delta^{ab}~.}} Compact hyperhermitian four-manifolds have been classified in~\Boyertwo. They are globally conformal to one of the following: a flat torus~$T^4$, a~$K3$ surface with Ricci-flat K\"ahler metric, or~$S^3 \times S^1$ with standard metric~$ds^2 = d\tau^2 + r^2 d \Omega_3$ and certain quotients thereof. However, the third option can be ruled out because these manifolds do not admit a complex structure whose anti-canonical bundle admits a nowhere vanishing, anti-holomorphic section. As we saw in section~3, this is necessary for the existence of any solution of the form~$(\zeta, 0)$.

\subsec{Two Supercharges of the Form~$(\zeta, 0)$ and~$(0, \t \zeta)$}

Here we study manifolds that admit two solutions~$(\zeta, 0)$ and~$(0, \t \zeta)$ of the equations~\speq. This is only possible if the background fields~$M$ and~$\t M$ vanish,
\eqn\zerom{M=0~,\qquad \t M=0~,}
so that the two equations in~\speq\ are independent. Since these equations are linear and homogenous, it follows that~$(\zeta, \t \zeta)$ is also a solution. Conversely, any solution~$(\zeta, \t \zeta)$ that satisfies~\speq\ and the additional condition~\zerom\ splits into two independent solutions~$(\zeta, 0)$ and~$(0, \t \zeta)$. Thus, we are simply studying solutions of the kind considered in subsection~3.2, subject to the additional requirement~\zerom.

This condition further constrains the metric~\ellmet\ and the section~$p$ of the anti-canonical bundle that determines the background fields~$b^\mu, M, \t M$ and the spinors. Here we will analyze these constraints locally, using the~$w, z$ coordinates of subsection~3.2. In these coordinates, the fields~$M$ and~$\t M$ are given by~\mmbcoord, and hence we impose
\eqn\mmbcoordbis{M = {2 i p \over \Omega^4 c^2}\,  \d_z \log p = 0~, \qquad \t M = - { 2i \Omega^2 \over p} \left(\d_{\b z} \log\bigg({\Omega^6 c^2 \over p }\bigg) + \b h \, \d_{\b w} \log p \right) = 0~.}
Recalling that~$p$ does not depend on~$w$, the first equation implies that~$p = p(\b w, \b z)$ is anti-holomorphic. As in~\pah, this also follows directly from the existence of the solution~$(\zeta, 0)$. The second equation in~\mmbcoordbis\ implies that
\eqn\seccon{\d_{\b z} \log(\Omega^6 c^2 p^{-1}) + \b h \, \d_{\b w} \log p =0~.}
Taking an additional~$\b w$-derivative,
\eqn\thcond{\b h \d^2_{\b w} \log p -\d_{\b z}\d_{\b w} \log p=0~.}
Unless~$h(z, \b z)$ is holomorphic,\foot{If~$h(z, \b z) = h(z)$ is holomorphic, we can set it to zero by a holomorphic coordinate change of the form~$w \rightarrow w + F(z)$. It follows from~\thcond\ that~$p(\b w, \b z) = A(\b w) B(\b z)$. Substituting into~\seccon, we conclude that~$\Omega^6 c^2$ can be set to~$1$ by a holomorphic coordinate change of the form~$z \rightarrow G(z)$.} the only quantity in~\thcond\ that depends on~$z$ is~$\b h$, so that
\eqn\solf{p(\b w, \b z) = e^{\alpha \b w} \hat p(\bar z)~,\qquad \alpha \in \C~.}
Here~$\hat p (\b z)$ is nowhere vanishing, and hence we can set~$\hat p(\b z) = 1$ by a holomorphic coordinate change of the form~$z \rightarrow G(z)$. If~$\alpha\neq 0$, we can use~\seccon\ to solve for~$h$ in terms of~$\Omega$ and~$c$,
\eqn\solhf{h=-{\b \alpha}^{\, -1} \d_z \log \Omega^6 c^2~.}
If~$\alpha = 0$, then~$p = 1$ and~\seccon\ implies that~$\Omega^6 c^2$ is a constant while~$h$ is undetermined.

With these constraints on the metric and~$p = e^{\alpha \b w}$, the two solutions~$(\zeta, 0)$ and~$(0, \t \zeta)$ are given by~\zztell,
\eqn\zztell{\zeta_\alpha = {\sqrt s \over 2} \pmatrix{0 \cr 1}~, \qquad \t \zeta^\alphadot = {\Omega \over \sqrt s} \pmatrix{0 \cr 1}~, \qquad s = e^{\alpha \b w} g^{- {1 \over 4}}~.}
The background field~$b^\mu$ is given by~\bc\ and \bccomp. The metric and~$b^\mu$ are invariant under the Killing vector~$K^\mu = \zeta \sigma^\mu \t \zeta$ and its complex conjugate,  so that both~$K$ and~$\b K$ are part of the supersymmetry algebra. If we denote the supercharges by~$Q = (\zeta, 0)$ and~$\t Q = (0, \t \zeta)$, we obtain the following (anti-)~commutation relations:
\eqn\supalgco{ \{\delta_Q, \delta_{\t Q}\}=2 i \delta_K~, \qquad [\delta_{\b K}, \delta_Q]=-{\alpha\over 2} \delta_Q, \qquad [\delta_{\b K}, \delta_{\t Q}]= {\alpha\over 2} \delta_{\t Q}~.}

\newsec{Manifolds Admitting Four Supercharges}

In this section we briefly discuss necessary conditions for the existence of four supercharges. A more detailed treatment can be found in~\FestucciaWS\ and references therein. Assuming the existence of four independent solutions of~\speq, we can use~\intcond\ to obtain the following integrability conditions:
\medskip
\item{1.)} The Weyl tensor vanishes, $W_{\mu\nu\rho\lambda} = 0$, so that~$\CM$ is locally conformally flat.
\item{2.)} The one-form~$b_\mu$ is covariantly constant, $\grad_\mu b_\nu = 0$.

\item{3.)} The scalars~$M$ and~$\t M$ are constant, $\d_\mu M = \d_\mu \t M = 0$.

\item{4.)} Either~$b_\mu = 0$ or~$M = \t M = 0$.

\item{5.)} The Ricci tensor is given by
\eqn\riccfsc{R_{\mu\nu} = {1 \over 3} g_{\mu\nu} M \t M - {2 \over 9} (b_\mu b_\nu - g_{\mu\nu} b^\rho b_\rho)~.}
\medskip
\noindent Therefore, the solutions fall into two classes:
\medskip
\item{(I)} If~$M = \t M = 0$, then~$\CM$ is locally isometric to~$\CM_3 \times \R$. It follows from~\riccfsc\ that~$\CM_3$ is a space of constant curvature and that~$b_\mu$ is either real or purely imaginary. This leads to the following three subcases:

\itemitem{(Ia)} If~$b^\mu b_\mu = - {1 \over r^2}$ with~$r>0$, then~$\CM_3$ is locally isometric to a round~$S^3$ of radius~$r$. In this case~$b_\mu$ is purely imaginary and points along~$\R$.

\itemitem{(Ib)} If~$b_\mu = 0$, then~$\CM$ is locally isometric to flat~$\R^4$. This is the case of ordinary flat-space supersymmetry.

\itemitem{(Ic)} If~$b^\mu b_\mu = {1 \over r^2}$ with~$r > 0$, then~$\CM_3$ is locally isometric to~$H^3$ of radius~$r$. In this case~$b_\mu$ is real and points along~$\R$.

\item{(II)} If~$b_\mu = 0$, it follows from~\riccfsc\ that~$M \t M$ is real, so that~$M$ and~$\t M$ must have opposite phase. Together with the fact that the Weyl tensor vanishes, this implies that~$\CM$ is a space of constant curvature. We then have the following three subcases:

\itemitem{(IIa)} If~$M \t M = - {9 \over r^2}$ with~$r > 0$, then~$\CM$ is locally isometric to a round~$S^4$ of radius~$r$.

\itemitem{(IIb)} If~$M \t M = 0$, then~$\CM$ is locally isometric to flat~$\R^4$. This is identical to case~(Ib) above.

\itemitem{(IIc)} If~$M \t M = {9 \over r^2}$ with~$r > 0$, then~$\CM$ is locally isometric to~$H^4$, the four-dimensional hyperbolic space of radius~$r$.

\vskip 1cm

\noindent {\bf Acknowledgments:}
We would like to thank N.~Seiberg for collaboration in the early stages of this project, and for many useful discussions. We are grateful to C.~Closset, Z.~Komargodski, and N.~Seiberg for comments on the manuscript. The work of TD was supported in part by a DOE Fellowship in High Energy Theory and a Centennial Fellowship from Princeton University. The work of GF was supported in part by NSF grant PHY-0969448 and a Marvin L. Goldberger Membership at the Institute for Advanced Study.  GF is grateful for the hospitality of the Aspen Center for Physics during the completion of this project. Any opinions, findings, and conclusions or recommendations expressed in this
material are those of the authors and do not necessarily reflect the views of the funding agencies.

\appendix{A}{Conventions}

We follow the conventions of~\WessCP, adapted to Euclidean signature. This leads to some differences in notation, which are summarized here, together with various relevant formulas.

\subsec{Flat Euclidean Space}

The metric is given by~$\delta_{\mu\nu}$, where~$\mu, \nu = 1, \ldots, 4$. The totally antisymmetric Levi-Civita symbol is normalized so that~$\ep_{1234} = 1$. The rotation group is given by~$SO(4) = SU(2)_+ \times SU(2)_-$. A left-handed spinor~$\zeta$ is an~$SU(2)_+$ doublet and carries un-dotted indices, $\zeta_\alpha$. Right-handed spinors~$\t \zeta$ are doublets under~$SU(2)_-$. They are distinguished by a tilde and carry dotted indices, $\t \zeta^\alphadot$. In Euclidean signature, $SU(2)_+$ and~$SU(2)_-$ are not related by complex conjugation, and hence~$\zeta$ and~$\tilde \zeta$ are independent spinors.

The Hermitian conjugate spinors~$\zeta^\dagger$ and~${\t \zeta}^{\dagger}$ transform as doublets under~$SU(2)_+$ and~$SU(2)_-$ respectively. They are defined with the following index structure,
\eqn\daggers{(\zeta^\dagger)^\alpha = \overline {(\zeta_\alpha)}~, \qquad (\t \zeta^\dagger)_\alphadot = \overline {(\t \zeta^\alphadot)}~,}
where the bars denote complex conjugation. Changing the index placement on both sides of these equations leads to a relative minus sign,
\eqn\daggersii{(\zeta^\dagger)_\alpha = - \overline{(\zeta^\alpha)}~, \qquad (\t \zeta^\dagger)^\alphadot = - \overline{(\t \zeta_\alphadot)}~.}
We can therefore write the~$SU(2)_+$ invariant inner product of~$\zeta$ and~$\eta$ as~$\zeta^\dagger \eta$. Similarly, the~$SU(2)_-$ invariant inner product of~$\t \zeta$ and~$\t \eta$ is given by~${\t \zeta}^\dagger \t \eta$. The corresponding norms are denoted by~$|\zeta|^2 = \zeta^\dagger \zeta$ and~$|\t \zeta|^2 = \t \zeta^\dagger \t \zeta$.

The sigma matrices take the form
\eqn\sigmamat{\sigma^\mu_{\alpha\alphadot} = (\vector{\sigma}, -i)~,\qquad \t \sigma^{\mu \alphadot\alpha} = (-\vector{\sigma}, -i)~,}
where~$\vector{\sigma} = (\sigma^1, \sigma^2, \sigma^3)$ are the Pauli matrices. We use a tilde (rather than a bar) to emphasize that~$\sigma^\mu$ and~$\t \sigma^\mu$ are not related by complex conjugation in Euclidean signature. The sigma matrices~\sigmamat\ satisfy the identities
\eqn\antic{\sigma_\mu\t \sigma_\nu + \sigma_\nu \t \sigma_\mu = -2\delta_{\mu\nu}~, \qquad \t \sigma_\mu \sigma_\nu + \t \sigma_\nu \sigma_\mu = -2\delta_{\mu\nu}~.}
The generators of~$SU(2)_+$ and~$SU(2)_-$ are given by the antisymmetric matrices
\eqn\smunu{\sigma_{\mu\nu} = {1 \over 4} (\sigma_\mu \t \sigma_\nu - \sigma_\nu\t \sigma_\mu)~, \qquad \t \sigma_{\mu\nu} = {1 \over 4} (\t \sigma_\mu \sigma_\nu - \t \sigma_\nu \sigma_\mu)~.}
They are self-dual and anti-self-dual respectively,
\eqn\asd{\half \ep_{\mu\nu\rho\lambda} \sigma^{\rho \lambda } = \sigma_{\mu\nu}~, \qquad \half \ep_{\mu\nu\rho\lambda} \t \sigma^{\rho\lambda} = - \t \sigma_{\mu\nu}~.}

\subsec{Differential Geometry}

We will use lowercase Greek letters~$\mu, \nu, \ldots$ to denote curved indices and lowercase Latin letters~$a, b, \ldots$ to denote frame indices. Given a Riemannian metric~$g_{\mu\nu}$, we can define an orthonormal tetrad~${e^a}_\mu$. The Levi-Civita connection is denoted~$\grad_\mu$ and the corresponding spin connection is given by
\eqn\spincon{{\omega_{\mu a}}^b = {e^b}_\nu \grad_\mu {e_a}^\nu~.}
The Riemann tensor takes the form
\eqn\riemann{{R_{\mu\nu a}}^b = \d_\mu {\omega_{\nu a}}^b - \d_\nu {\omega_{\mu a}}^b + {\omega_{\nu a}}^c {\omega_{\mu c}}^b - {\omega_{\mu a}}^c {\omega_{\nu c}}^b~.}
The Ricci tensor is defined by~$R_{\mu\nu} = {R_{\mu\rho\nu}}^\rho$, and~$R = {R_\mu}^\mu$ is the Ricci scalar. Note that in these conventions, the Ricci scalar is negative on a round sphere.

The covariant derivatives of the spinors~$\zeta$ and~$\t \zeta$ are given by
\eqn\covsp{\grad_\mu \zeta = \d_\mu \zeta + \half \omega_{\mu a b } \sigma^{ab} \zeta~, \qquad \grad_\mu \t \zeta = \d_\mu \t \zeta + \half \omega_{\mu a b } {\t \sigma}^{ab} \t \zeta~.}
We will also need the commutator of two covariant derivatives,
\eqn\comspd{[\grad_\mu, \grad_\nu] \zeta = \half R_{\mu\nu a b} \sigma^{ab} \zeta~, \qquad [\grad_\mu, \grad_\nu] \t\zeta = \half R_{\mu\nu a b} {\t \sigma}^{ab} \t \zeta~.}
Finally, the Lie derivatives of~$\zeta$ and~$\t \zeta$ along a vector field~$X = X^\mu \d_\mu$ are given by~\refs{\Kosmann},
\eqn\lsp{\eqalign{& \CL_X \zeta=X^\mu \grad_\mu \zeta -{1\over 2} \grad_\mu X_\nu\sigma^{\mu\nu} \zeta~,\cr
&  \CL_X \t \zeta=X^\mu \grad_\mu \t \zeta -{1\over 2} \grad_\mu X_\nu{\t \sigma}^{\mu\nu} \t \zeta~.}}

\appendix{B}{The Chern Connection}

In this appendix we review some useful facts about the Chern connection on a Hermitian manifold. In general, a connection~$\hat \grad_\mu$ is metric compatible, $\hat \grad_\mu g_{\nu\rho} = 0$, if and only if its connection coefficients~${\, \hat \Gamma^\mu}_{\nu\rho}$ can be expressed as
\eqn\cont{\eqalign{& {{\, \hat \Gamma}^\mu}_{\nu\rho} = {{\Gamma}^\mu}_{\nu\rho} + {K^\mu}_{\nu\rho}~, \cr
& {{\Gamma}^\mu}_{\nu\rho} = \half g^{\mu \lambda} \left(\d_\nu g_{\rho\lambda} + \d_\rho g_{\nu\lambda} - \d_\lambda g_{\nu\rho}\right)~, \cr
 & K_{\mu\nu\rho} = - K_{\rho\nu\mu}~.}}
Here~${K^\mu}_{\nu\rho}$ is the contorsion tensor. If we set it to zero, we recover the usual Levi-Civita connection~$\grad_\mu$. The spin connection corresponding to~$\hat \grad_\mu$ is given by
\eqn\sc{\hat \omega_{\mu\nu\rho} = \omega_{\mu\nu\rho} -K_{\nu\mu\rho}~,}
where~$\omega_{\mu\nu\rho}$ is the spin connection associated with the Levi-Civita connection.

Given an almost complex structure~${J^\mu}_\nu$, the Nijenhuis tensor is defined by
\eqn\njdef{{N^\mu}_{\nu\rho} = {J^\lambda}_\nu \grad_\lambda {J^\mu}_\rho - {J^\lambda}_\rho \grad_\lambda {J^\mu}_\nu -{J^\mu}_\lambda \grad_\nu {J^\lambda}_\rho + {J^\mu}_\lambda \grad_\rho {J^\lambda}_\nu~.}
If~${J^\mu}_\nu$ is also compatible with the metric, $g_{\mu\nu} {J^\mu}_\alpha {J^\nu}_\beta = g_{\alpha\beta}$, it is straightforward to verify the following identity:
\eqn\magicid{J_{\mu\lambda} {N^\lambda}_{\nu\rho} = 2 \grad_\mu J_{\nu\rho} + {J_\nu}^\alpha {J_\rho}^\beta (dJ)_{\mu\alpha\beta} - (dJ)_{\mu\nu\rho}~,}
where~$(dJ)_{\mu\nu\rho} = \grad_\mu J_{\nu\rho} + \grad_\nu J_{\rho\mu} + \grad_\rho J_{\mu\nu}$. This formula is especially useful when the complex structure is integrable, so that~${N^\lambda}_{\nu\rho} = 0$. In this case it expresses the covariant derivative~$\grad_\mu J_{\nu\rho}$ in terms of the exterior derivative~$(dJ)_{\mu\nu\rho}$.

Given an integrable complex structure~${J^\mu}_\nu$ and a compatible Hermitian metric, we define the Chern connection~$\grad^c_\mu$ as a metric compatible connection with contorsion tensor
\eqn\chern{K_{\nu\mu\rho} = \half {J_\mu}^\lambda (dJ)_{\lambda \nu\rho}~.}
It follows from~\magicid\ and the fact that~${N^\lambda}_{\nu\rho} = 0$ that~$\grad^c_\mu J_{\nu\rho} = 0$, so that the Chern connection is also compatible with the complex structure.

We are interested in complex structures that are given in terms of a spinor~$\zeta$ as in~\acs. In this case the derivative of~$\zeta$ with respect to the associated Chern connection is given by
\eqn\dcz{\grad_\mu^c \zeta = \grad_\mu \zeta - {i \over 4} \grad_\rho J^{\rho\nu}(g_{\mu\nu} + i J_{\mu\nu})  \zeta - {i \over 2} \grad_\rho J^{\rho\nu} \sigma_{\mu\nu}\zeta~.}

\appendix{C}{Two Supercharges on Warped~$T^3 \times \R$ and Warped~$H^3 \times \R$}

In this appendix, we show that two supercharges~$(\zeta, \t \zeta)$ and~$(\eta, \t \eta)$, both of which square to a non-trivial Killing vector that commutes with its complex conjugate, can only exist on a warped product~$T^3 \times \R$ or a warped product~$H^3 \times \R$.

Since~$(\zeta, \t \zeta)$ and~$(\eta, \t \eta)$ both satisfy~\speq, we can use the spinors to construct three Killing vectors,
\eqn\tkv{K^\mu = \zeta \sigma^\mu \t \zeta~, \qquad K'^\mu = \eta \sigma^\mu \t \eta~, \qquad Y^\mu = \zeta \sigma^\mu \t \eta + \eta \sigma^\mu \t \zeta~,}
and a conformal Killing vector,
\eqn\ckv{C^\mu = \zeta \sigma^\mu \t \eta - \eta \sigma^\mu \t \zeta~.}
Similarly, we can use~\speq\ to compute several useful commutators,
\eqn\usefulcomms{\eqalign{&[K,K']=\alpha Y-{i\over 3}(b^\mu Y_\mu) C~, \qquad \alpha=-{i \over 3} \big(M \t\zeta\t\eta + \t M \zeta\eta + b^\mu C_\mu\big)~,\cr
&[K,Y] = 2 \alpha K-{2i\over 3}(b^\mu K_\mu) C~,\cr
&[K,C]= {2i\over 3}(b^\mu Y_\mu) K -{2i\over 3}(b^\mu K_\mu) Y~.}}
Since the conformal Killing vector~$C$ cannot appear in the commutators of the Killing vectors~$K, K'$, and~$Y$, we conclude that
\eqn\dotzero{b^\mu K_\mu = b^\mu Y_\mu = 0~, \qquad \alpha = {\rm constant}~.}
This follows from the fact that~$K, K'$, and~$Y$ must form a closed algebra (otherwise we could commute any additional Killing vectors with~$(\zeta, \t \zeta)$ and~$(\eta, \t \eta)$ to obtain additional supercharges), and the fact that~$C$ is linearly independent of~$K, K'$, and~$Y$ at every point (otherwise~$(\zeta, \t \zeta)$ and~$(\eta, \t \eta)$ would not be independent solutions).

We will work with the linear combinations
\eqn\lincomb{S^\mu = \half (Y^\mu + C^\mu) = \zeta \sigma^\mu \t \eta~, \qquad \t S^\mu = \half (Y^\mu - C^\mu) = \eta \sigma^\mu \t \zeta~,}
both of which are conformal Killing vectors. Note that~$S^\mu$ is holomorphic with respect to the complex structure~${J^\mu}_\nu$ constructed from~$\zeta$, while~$\t S^\mu$ is holomorphic with respect to the complex structure~${\, \t J^\mu}_\nu$ constructed from~$\t \zeta$,
\eqn\holceq{{J^\mu}_\nu S^\nu = i S^\mu~, \qquad {\, \t J^\mu}_\nu \t S^\nu = i \t S^\mu~.}
As in subsection~3.2, we can use the complex structure~${J^\mu}_\nu$ to introduce holomorphic~$w, z$ coordinates, so that the metric takes the form~\ellmet,
\eqn\ellmetbis{ds^2 =\Omega(z,\bar z)^2\left( (dw +h(z,\bar z) dz)(d\bar w +\b h(z,\bar z) d\bar z)+ c(z,\bar z)^2 dz d\bar z\right)~,}
with~$K = \d_w$. Since~$(\zeta, \t \zeta)$ is a supercharge of the from considered there, it follows from the expressions~\bc,~\bccomp\ for~$b_\mu$ and the requirement~\dotzero\ that
\eqn\hholc{K^\mu b_\mu = -{3 i \over 2 c^2} \d_z \b h = 0~.}
Therefore~$h(z, \b z) = h(z)$ is holomorphic and can be set to zero by a holomorphic coordinate change of the form~$w \rightarrow w + F(z)$, so that the metric~\ellmetbis\ reduces to
\eqn\metsimple{ds^2 =\Omega(z,\bar z)^2\left(dw d\bar w + c(z,\bar z)^2 dz d\bar z\right)~.}
It is now straightforward to switch between the complex structures~${J^\mu}_\nu$ and~${\, \t J^\mu}_\nu$ by simply exchanging~$z$ and~$\b z$.

We can constrain the form of~$S^\mu$ by using the fact that it is holomorphic with respect to~${J^\mu}_\nu$, and by imposing the conformal Killing equation and the commutation relation that follows from~\usefulcomms\ and~\dotzero,
\eqn\sfmla{\grad_\mu S_\nu + \grad_\nu S_\mu = A g_{\mu\nu}~, \qquad [K, S] = \alpha K~,}
where~$A$ is a scalar function. It follows that the real function~$c(z, \b z)$ in~\metsimple\ is the product of a holomorphic and an anti-holomorphic function, so that we can set to~$c(z, \b z) = 1$ by a holomorphic coordinate change of the form~$z \rightarrow  G(z)$. The metric now only depends on the real function~$\Omega(z, \b z)$. Moreover, $S^\mu$ and~$A$ must take the form
\eqn\scomp{\eqalign{& S = (\alpha w + \beta \b z + \gamma)\d_w + (\alpha z - \beta \b w + \delta) \d_z~, \qquad~\beta, \gamma, \delta \in \C~,\cr
& A = \alpha + (\alpha z - \beta \b w + \delta) \d_z \log \Omega^2~.}}

We can repeat the same analysis for the conformal Killing vector~$\t S^\mu$, which is holomorphic with respect to~${\, \t J^\mu}_\nu$ and satisfies
\eqn\stfmla{\grad_\mu \t S_\nu + \grad_\nu \t S_\mu = -A g_{\mu\nu}~, \qquad [K, \t S] = \alpha K~.}
The appearance of~$-A$ in the first equation follows from the fact that~$S + \t S = Y$ is a genuine Killing vector. Therefore,
\eqn\stcomp{\eqalign{& \t S = (\alpha w + \t \beta z + \t \gamma) \d_w + (\alpha \b z - \t \beta \b w + \t \delta) \d_{\b z}~, \qquad \t \beta, \t \gamma, \t \delta \in \C~,\cr
& A = - \alpha - (\alpha \b z - \t \beta \b w + \t \delta) \d_{\b z} \log \Omega^2~.}}

Comparing~\scomp\ and~\stcomp, we conclude that
\eqn\conscond{\eqalign{& (\beta \d_z - \t \beta \d_{\b z}) \Omega(z, \b z) = 0~,\cr
& \big( (\alpha z + \delta) \d_z  + (\alpha \b z + \t \delta) \d_{\b z}  +  \alpha\big)\Omega(z, \b z)= 0~.}}
If~$\alpha = 0$, then~$\Omega$ is annihilated by~$\beta \d_z - \t \beta \d_{\b z}$ and~$\delta \d_z + \t \delta \d_{\b z}$. Unless~$\Omega$ is a constant, these two linear combinations must be proportional to the same real vector. By using the remaining coordinate freedom to redefine the phase of~$z$, we can thus choose~$\Omega$ to only depend on the real part of~$z$. The metric is then given by
\eqn\tormet{ds^2 = \Omega(\Re z)^2 (dw d \b w + dz d \b z)~.}
Defining a real coordinate~$\tau$ via~$d\tau = \Omega(\Re z) d (\Re z)$, the metric~\tormet\ takes the form~$ds^2 = d\tau^2 + \Omega(\tau)^2 (dw d\b w + d(\Im z)^2)$, which describes warped~$T^3 \times \R$.

If~$\alpha \neq 0$ and~$\beta, \t \beta, \delta, \t \delta$ vanish, the constraints~\conscond\ imply that~$\Omega$ must take the form
\eqn\omegasol{\Omega(z, \b z) = {1 \over |z|} \hat \Omega\left({i \over 2} \log{z \over \b z}\right)~.}
Here the positive function~$\hat \Omega$ only depends on the phase of~$z$. Introducing a new holomorphic coordinate~$u = i \log z$, we obtain the metric
\eqn\warphtmet{ds^2 = \hat \Omega(\Re u)^2 (e^{-2 \Im u} dw d\b w + d u d\b u)~.}
In terms of the real coordinates~$\tau, x_{1,2,3}$ defined by~$d \tau = \hat \Omega(\Re u) d(\Re u)$, $x_1 = e^{\Im u}$, and~$w = x_2 + i x_3$, the metric takes the form
\eqn\finalmet{ds^2 = d\tau^2 + \hat \Omega(\tau)^2 \; {dx_1^2 + dx_2^2 + dx_3^2 \over x_1^2}~,}
and hence it describes warped~$H^3 \times \R$. As in the warped~$T^3 \times \R$ case above, the presence of the constants~$\beta, \t \beta, \delta, \t \delta$ in~\conscond\ may further constrain the function~$\Omega(z, \b z)$ in~\omegasol\ to only depend on the real part of~$z$. (It may also shift the origin of~$z$.) Up to a constant factor, this fixes~$\Omega \sim {1 \over \Re z}$, so that the metric~$ds^2 \sim {1 \over (\Re z)^2} (dw d\b w + dz d \b z)$ describes~$H^4$.

\listrefs

\end